\documentclass[%
superscriptaddress,
preprint,
showpacs,
 amsmath,amssymb,
 aps,
sd,
]{revtex4-1}

\usepackage{ulem}
\usepackage{xcolor}
\usepackage{graphicx}% Include figure files
\usepackage{dcolumn}% Align table columns on decimal point
\usepackage{bm}% bold math
\usepackage{nicefrac}
\usepackage{units}
\usepackage{textcomp}
\usepackage [autostyle, english = american]{csquotes}
\MakeOuterQuote{"}
\usepackage[]{natbib} % Use the natbib reference package - read up on this to edit the reference style; if you want text (e.g. Smith et al., 2012) for the in-text references (instead of numbers), remove 'numbers'

\begin{document}

\preprint{APS/123-QED}

\title[Hydrodynamic pairing of {soft particles} in a confined flow]{Hydrodynamic pairing of {soft particles} in a confined flow}% Force line breaks with \\
%\thanks{Footnote to title of article.}

\author{O. Aouane}%
\altaffiliation[Current address: ]{Forschungszentrum J\"{u}lich GmbH, Helmholtz-Institute Erlangen-N\"{u}rnberg for Renewable Energy (IEK-11), Dynamics of Complex Fluids and Interfaces, F\"{u}rther Stra\ss{}e 248, 90429 N\"{u}rnberg, Germany
 }%
 \affiliation{
 Universit\'{e} Grenoble Alpes, LIPHY, F-38000, Grenoble, France
 }%
 \affiliation{
 CNRS, LIPHY, F-38000, Grenoble, France
 }%
 \affiliation{
 Experimental Physics, Saarland University, 66123 Saarbr\"{u}cken, Germany.%\\This line break forced with \textbackslash\textbackslash
 }%
\author{A. Farutin}%
 \affiliation{
 Universit\'{e} Grenoble Alpes, LIPHY, F-38000, Grenoble, France
 }%
 \affiliation{
 CNRS, LIPHY, F-38000, Grenoble, France
 }%
\author{M. Thi\'{e}baud}%
 \affiliation{
 Universit\'{e} Grenoble Alpes, LIPHY, F-38000, Grenoble, France
 }%
 \affiliation{
 CNRS, LIPHY, F-38000, Grenoble, France
 }%
\author{A. Benyoussef}%
 \affiliation{
 LMPHE, URAC 12, Facult\'{e} des Sciences, Universit\'{e} Mohammed V- Agdal, Rabat, Morocco%\\This line break forced with \textbackslash\textbackslash
 }%
\author{C. Wagner}%
 \affiliation{
 Experimental Physics, Saarland University, 66123 Saarbr\"{u}cken, Germany.%\\This line break forced with \textbackslash\textbackslash
}%
\author{C. Misbah}
 \affiliation{
 Universit\'{e} Grenoble Alpes, LIPHY, F-38000, Grenoble, France
 }%
 \affiliation{
 CNRS, LIPHY, F-38000, Grenoble, France
 }%

\date{\today}% It is always \today, today,

\begin{abstract}
{The mechanism of hydrodynamics-induced pairing of soft particles, namely closed bilayer membranes (vesicles, a model system for red blood cells) and drops,} is studied numerically with %
a special attention paid to the role of the confinement (the particles are within two rigid walls). This study unveils the complexity of the pairing mechanism due to hydrodynamic interactions. %
{We find both for vesicles and for drops that two particles attract each other and form a stable pair at weak confinement if their initial separation is below a certain value.} %
If the initial separation is beyond that distance, the particles repel each other and adopt a longer stable interdistance. %
This means that for the same confinement we have (at least) two stable branches. {To which branch a pair of particles relaxes with time depends only on the initial configuration. }%
An unstable branch is found between these two stable branches. At a critical confinement the stable branch corresponding to  the shortest interdistance merges %
with the unstable branch in the form of a saddle-node bifurcation. At  this critical confinement we have a finite jump from a solution corresponding to the continuation of the unbounded case to a solution which is induced by the presence of walls.
{The results are summarized in a phase diagram, which proves to be of a complex nature. The fact that both vesicles and drops have the same qualitative phase diagram points to the existence of a universal behavior, highlighting the fact that with regard to pairing the details of mechanical properties of the deformable particles are unimportant. This offers an interesting perspective for simple analytical modeling.}

% 47.11.Hj: Computational Methods in fluid dynamics
% 47.15.G-: Low-Reynolds number flows
% 83.50.Ha: Flow in channels
% 83.80.Lz: Physiological materials (e.g. blood, collagen, etc.)
% 87.16.D-: Membranes, bilayers, and vesicles
\end{abstract}

%Use showkeys class option if keyword
\maketitle

\section{\label{sec:intro}Introduction:\protect}
In the microcirculation, it is often observed that the red blood cells (RBCs) flow in single or multiple files forming small trains of cells, called clusters\cite{skalak69,gaehtgens82}. %
The arrangement and organization of the RBCs depend on the diameter of the vessel and their concentration  (hematocrit). %
Each RBC interacts hydrodynamically with the other cells\cite{tomaiuolo2012red}. %
RBCs can also interact via another mechanism, namely an interaction mediated by plasma proteins. The latter interaction is  materialized either by  bridging between RBCs or by a depletion force. %
In the bridging mechanisms proteins make a real bridge between two neighboring RBCs, while in the depletion mechanism osmosis is responsible for the cluster formation. %
We have recently discussed the implication of plasma proteins in the formation of RBC clusters in microcirculation \cite{brust2014plasma}. %
The main objective is to gain further insight into the role of each mechanism. Therefore this paper is be directed towards numerical {study of the effect of hydrodynamic interactions on cluster formation}. %

Several studies have been devoted  to understanding  the hydrodynamical interaction between suspended particles in the Stokes regime. Analytical models \cite{wang1969,leichtberg1976stokes} %
considered the motion of a linear array of rigid spheres at low Reynolds number in a cylindrical tube under a pressure-driven flow (i.e., an imposed Poiseuille flow).
Wang and Skalak\cite{wang1969} estimated {the range of the hydrodynamic interactions }between the spherical particles to be of the order of the tube diameter. %
Leichtberg \textit{et al.}\cite{leichtberg1976stokes} showed that the interparticle interactions were relatively small at weak confinements, reached a maximum at intermediate confinement, %
and were quickly damped out at strong confinement. More recently, colloidal particles confined between two parallel plates in a quasi 2D geometry have been studied experimentally and theoretically~\cite{Cui2002,Cui2004,Diamant2005}. %
The complexity in these systems arises from the difficulty to decouple the effect of Brownian diffusion from hydrodynamic interactions. %
An antidrag between the moving particles attributed to a negative hydrodynamic coupling has been reported. A change of sign of the hydrodynamic coupling (from attraction to repulsion) in a cylindrical channel was also observed \cite{Cui2002}. %
The effect of boundaries on the hydrodynamic interactions has been studied in the case of water-in-oil drops in quasi-1D microfluidic devices having a square section of the order of the size of the drop, so that the drops are constrained to move along the channel axis \cite{beatus2006phonons,beatus2007anomalous,shani2014long}. %
This study reported on a non-monotonous behavior of the hydrodynamic interaction resulting from an interplay between the plug flow and the screening of the long-range hydrodynamic interaction induced by the confinement. \\
Janssen \textit{et al.}\cite{janssen2012collective} have studied numerically pairs of rigid spheres and deformable drops driven by a Poiseuille flow through a three-dimensional (3D) rectangular channel in the Stokes regime. %
Due to the reversibility of Stokes equations, the interdistance between  a pair of rigid spheres does not evolve in time. However, for a pair of deformable drops, due to the up-stream/down-stream shape asymmetry, %
hydrodynamic interaction leads to an attraction at long interdistances and a repulsion at short interdistances.
The long-range attraction was attributed to the source-quadrupole flows induced by drop in the Hele-Shaw geometry.
%The repulsive nature has been attributed to the short-range dipolar interactions whereas the deformability of the drops induces a %far-field quadrupolar  attraction. %
The pair of drops tends to the same stationary interdistance independently of the capillary number, a measure of the flow strength, which only affects the time needed to reach the steady state. %
%\sout{Vesicles, a closed bilayer membrane of phospholipid molecules, have been used as a simplified model for RBCs to simulate blood flow in microcirculation \cite{mcwhirter2009flow,mcwhirter2011deformation} with a focus on the role of the imposed flow velocity and the hematocrit (concentration of cells in the tube) in the organization of cells in a capillary. The precise role of the interplay between confinement and the cell deformability is still not fully elucidated}.

{The rheological behavior and spatial organization of a suspension of vesicles and capsules in 2D and 3D under shear and parabolic flows have been widely studied numerically by several groups   \cite{breyiannis2000simple,liu2006rheology,secomb2007two,dupin2007modeling,vlahovska2009vesicles,doddi2009three,mcwhirter2009flow,veerapaneni2009boundary,zhao2010spectral,zhao2011shear,mcwhirter2011deformation,freund2011cellular,kruger2011particle,fedosov2011predicting,alizadehrad2012quantification,reasor2012coupling,zhao2013dynamics,marine2013wallsgreenfunction,freund2014numerical,matsunaga2015rheology,bryngelson2016capsule}.
Still, a complete determination of the phase diagram of the paired states has not yet been achieved.
For example, is the branch of the stationary solutions (say the stationary interdistance of a pair as a function of the confinement) unique, or are there many branches? If many branches exist, is there a coexistence domain, and how the topology of the bifurcation diagram evolves with parameters. The present study is focused along this question and reveals a phase diagram of a complex nature.}%

It has been shown in a recent study that stable clusters of vesicles can form in the absence of  bounding walls under an imposed parabolic flow profile \cite{Giovanni2012}.
Our objective in this paper is to take this unbounded case as a reference and see how confinement affects the hydrodynamic interactions.
We generically find repulsion at short interdistance and attraction at long interdistance. %
We also find that at a long enough  vesicle  interdistance the interaction can change sign, become repulsive,  and then become attractive at longer interdistance. %
{This points to the existence of an interaction that changes sign  with distance, highlighting the nontrivial effect of hydrodynamic interactions. %
We further analyze the complex structure of the branches of coexisting stationary solutions.  We mainly focus this study on vesicles, as a closer biomimetic counterpart for RBCs, but we also investigate the hydrodynamic interactions between drops in order to investigate whether or not there is a generic pattern behind our findings and assess how sensitive the hydrodynamic interactions are to the mechanical properties of the particles. We show that the bifurcation diagram for drops is similar to that for vesicles (both qualitatively and almost quantitatively), pointing to an underlying universality.}

A systematic 2D numerical study (based on a boundary integral formulation) is undertaken here  in order to analyze  the time evolution of a pair of vesicles {or a pair of drops} in a pressure-driven flow, %
by exploring several parameters, such as channel width, initial separation of the particles, and the flow strength. %
%We then consider the case of a larger clusters of vesicles and show that the size of a stable cluster can be drastically modified by the presence of walls. Implication of the results to real situations is discussed. %
This paper is organized as follows.  In Section \ref{sec:model} we introduce the model in detail. Section \ref{sec:results} presents the main results {for vesicles}, both for weak and strong confinement. {Section \ref{sec:drops} gives an overview of the results for drops. }%
Section \ref{sec:conclusion} is devoted to a discussion of the results and the implication for real situations.
\section{\label{sec:model}The model and the solution method}
{We first present  the method adopted in this study to solve the motion of two hydrodynamically interacting vesicles or drops confined between two parallel plates and subject to a Poiseuille flow. We start by introducing the theoretical model and then discuss the numerical scheme and the precision.} %
\subsection{\label{sec:model_sublevel1} {Mechanical model for vesicles and drops}}
Vesicles, drops, capsules (drops coated with polymers) are endowed with different mechanical properties leading to different responses to external stresses. %
 Both vesicles\cite{misbah2006,noguchi2007swinging,farutin2012vesicle,Othmane2014} and capsules\cite{ramanujan1998deformation,lac2005deformation,kessler2008swinging} are widely used to mimic RBCs under  flow. %
RBC's complex dynamics, such as tank-treading, tumbling and vacillating breathing (aka swinging or trembling) can be reproduced by both vesicles and capsules. %
Shapes exhibited by RBCs under a Poiseuille flow, such as parachute and slipper shapes, are also captured by the vesicle and capsule models (see review \cite{vlahovska2009vesicles}). %
Vesicles resist bending, and possess  an inextensible membrane (constant area in 3D and perimeter in 2D) but do not present a resistance to shearing, in contrast to RBCs and capsules, which are endowed with surface shear elasticity. %
Nevertheless, several features exhibited by RBCs are also captured by vesicles, such as the above-mentioned dynamics and shapes under flow. %

{The membrane energy of vesicles \cite{helfrich1973}} is expressed in its 2D form as
\begin{equation}
E_c = \frac{\kappa}{2} \oint{c^2 ds}+ \oint{\zeta ds},
\label{eq::helfrich}
\end{equation}
where $\kappa$ is the membrane bending modulus, $c$ is the  curvature, $ds$ is the arclength element, and $\zeta$ is a Lagrange multiplier, which enforces membrane  inextensibility.

{Liquid drops resist deformation by surface tension (line tension in 2D) forces, which tend to restore their shape to a sphere (a circle in 2D).
The tension energy is written as 
\begin{equation}
E_t = \zeta_0\oint{ds},
\label{eq::surfacetension}
\end{equation}
where $\zeta_0$ is the surface tension. We can see, that eq. (\ref{eq::helfrich}) is reduced to (\ref{eq::surfacetension}) by setting $\kappa=0$ and by treating $\zeta=\zeta_0$ as a constant representing the surface tension, which is an intrinsic quantity for drops (while for vesicles, $\zeta$ is not intrinsic to the membrane, but is an auxiliary field to enforce a constant local arclength and thus can change according to membrane load).}

The 2D membrane force is  obtained by calculating the functional derivative of the energy (\ref{eq::helfrich})
\begin{equation}
\mathbf{f} = 
\begin{cases}
\kappa\left(\frac{\partial^2{c}}{\partial s^2} + \frac{c^3}{2}\right)\mathbf{n}  - c \zeta \mathbf{n} + \frac{\partial{\zeta}}{\partial{s}}\mathbf{t}\quad \text{for vesicles} \\
-c\zeta_0\mathbf{n} \quad \text{for drops},
\end{cases}
\label{eq::membrane_force}
\end{equation}
where $\mathbf{n}$ and $\mathbf{t}$ are the outward normal and the tangent unit vectors, respectively. The details of the derivation can be found in Ref. \cite{kaoui2008migration}. 

{Both for vesicles and for drops, the particle size $R_v,$ which serves as the length scale of the problem, is defined by the expression 
\begin{equation}
S = \pi R_v^{2},
\end{equation}
where $S$ is the area inside the particle contour. Here and below, we use generic term particle to denote both drops and vesicles when there is no need to make a distinction between them.
Unlike drops, vesicles (and RBCs) are characterized by their reduced area
\begin{equation}
\nu = \frac{S}{\pi[p/(2\pi)]^2},
\end{equation}
an intrinsic dimensionless parameter that expresses the ratio between the actual fluid area enclosed by the vesicle contour and the area of a disk having the same perimeter $p$ as the vesicle.
For RBCs, the reduced volume (3D equivalent of the reduced area) lies in the range of $0.60-0.65$. Accordingly, the reduced area $\nu$ is kept fixed to $
0.65$ in all our simulations.
The equilibrium shape obtained by minimizing (\ref{eq::helfrich}) for vesicles with reduced volume $\nu = 0.60-0.65$ is a %
biconcave shape similar to that exhibited by a RBC at rest \cite{canham1970minimum}.}%
%\subsection{\label{sec:model_sublevel2}Problem statement}
%\subsubsection{\label{sec:model_subsublevel2}Governing equations}
\subsection{\label{sec:model_sublevel2}Boundary integral formulation}
\begin{figure}[h!]		
	\centering
	\includegraphics[width = 0.8 \linewidth]{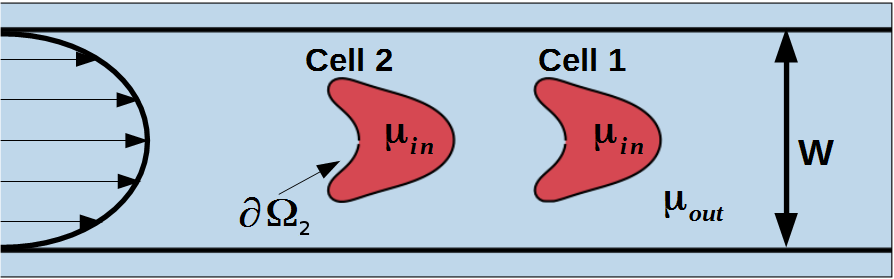}	
	\caption{Illustration of two vesicles driven by a Poiseuille flow and confined between two parallel walls situated at $x_2=\pm W/2$.}
	\label{fig::fig0}
\end{figure}

{We consider a system of two identical hydrodynamically interacting particles driven by a Poiseuille flow in a confined geometry between two laterally infinite plates (see Fig.\ref{fig::fig0}). The fluids inside and outside the particles have the same densities $\rho_{in}=\rho_{out}=\rho$, and the same dynamic viscosities $\mu_{in} = \mu_{out} = \mu$. %
The velocity in the absence of particles  (i.e. the undisturbed velocity) is  denoted by $\mathbf{u}^\infty$ and its Cartesian components are given by
\begin{equation}
\left\{
\begin{array}{rcr}
u_1^\infty (\boldsymbol {x}) & = & u_{max}\left[1-(\frac{x_2}{W/2})^2\right] \\
u_2^\infty (\boldsymbol {x}) & = & 0 
\end{array}
\right. 			
\end{equation}%
where $\boldsymbol {x}(x_1,x_2)$ is an arbitrary point in the whole domain, $u_{max}$ is the midplane velocity and $W$ is the channel width. $x_1$ and $x_2$ are the cartesian coordinates of $\boldsymbol {x}$ along the flow and in the perpendicular directions, respectively.} %
The Reynolds number associated with a vesicle (with typical size $10\mu m$, water viscosity and speed in the range of mm/s, as found in microcirculation) is of order $0.01$. The parameters for drops are chosen so that the Reynolds number remains small as compared to unity. Thus the motion of the fluids can safely be described by the Stokes equations
\begin{align}
\label{eq::motion_stokes}
&\nabla \cdot \sigma(\boldsymbol{x})  = - \nabla P(\boldsymbol{x})+ \mu \nabla^2 \mathbf{u}(\boldsymbol{x}) = 0\\
&\nabla \cdot \mathbf{u}(\boldsymbol{x})  = 0
\label{eq::continuity_stokes}
\end{align}
where $\mathbf{\sigma}$ is the stress tensor associated to the total velocity and $P$ is the pressure. The jump in the interfacial traction is balanced by the particle interface  force $\mathbf{f}= -[\sigma_{out} - \sigma_{in}] \mathbf{n}$. %
Besides the latter force balance condition, the solution of equations (\ref{eq::motion_stokes}) and (\ref{eq::continuity_stokes}) must respect the following boundary conditions (B.C.)
\begin{align}
\label{eq::bc1}
	&\mathbf{u}(\boldsymbol{x}) = \mathbf{0},  \quad \text{when} \hspace{.15cm} \boldsymbol{x} \hspace{.15cm} \text{lies on the walls } \\
\label{eq::bc2}	
	&\lim_{\boldsymbol{x}\to\infty} [\mathbf{u}(\boldsymbol{x}) - \mathbf{u}^\infty(\boldsymbol{x})] = \mathbf{0} \\
\label{eq::bc3}	
	&\mathbf{u}_{out}(\boldsymbol{x})=\mathbf{u}_{in}(\boldsymbol{x}), \quad \boldsymbol{x} \in \partial \Omega_i \quad (i=1,...,N_{v})
\end{align}
where the subscripts "in" and "out" refer to the fluid inside and outside the particle, $\partial \Omega_i$ is the contour of the $i$th particle and $N_v$ is the total number of particles. %   
{ The Stokes equations can be converted into a boundary integral equation \cite{pozrikidis1992boundary} defined over $\partial \Omega_i,$ {which reads}
\begin{equation}
  \mathbf{u}(\boldsymbol{x}) = \mathbf{u}^{\infty}(\boldsymbol{x}) %
  + \frac{1}{4\pi \mu} \sum_{i}\oint\limits_{{\partial \Omega_{i}}} {\mathbf{\mathcal{G}}}^{2W}(\boldsymbol{x},\boldsymbol{y}) \cdot \mathbf {f} (\boldsymbol{y}) ds(\boldsymbol{y}) 
  \label{eq::boundary_integral}
\end{equation}
where $\mathbf{\mathcal{G}}^{2W}$ is a Green's function satisfying the appropriate boundary conditions {(i.e. it vanishes at the walls and at infinity)},  $\boldsymbol {x}(x_1,x_2)$ and  $\boldsymbol {y}(y_1,y_2)$ are the target and source points, {and $ds$ is the element of membrane arclength.} The whole right hand side in (\ref{eq::boundary_integral}) is the total flow field. The integral on the right hand side is performed over the particles' contours $\partial \Omega_{i}$ and corresponds to  the contribution of the particles to the total velocity field.  This contribution is referred to as the {induced flow field}. We also focus on the flow field in the co-moving frame, that is the frame moving with the pair of vesicles (when they reach a steady-state interdistance).
\subsection{\label{sec:method_sublevel1}{Numerical method}}
\subsubsection{{Dimensionless form}}
{Most of the governing equations are the same for vesicles and drops. The difference appears only in the interfacial force expression and in the characteristic time of shape relaxation to equilibrium. Dimensionless quantities are denoted with a star symbol and are defined as
\begin{equation}
c^* = cR_{v},\,\,\,s^* = s/R_{v},\,\,\,\mathbf{u}^* = \mathbf{u} \tau_c/R_{v},\,\,\, \boldsymbol{x}^{*} = \boldsymbol{x}/R_{v}
\end{equation}
both for vesicles and for drops.
The shape relaxation time $\tau_c$ is taken as a time unit and is defined  as
\begin{equation} 	
\tau_c = 
\begin{cases}
\frac{\mu R_{v}^3}{\kappa} \quad  \text{for vesicles} \\
\frac{\mu R_{v}}{\zeta_0} \quad \text{for drops}.
\end{cases}
%&\tau_c = \frac{\mu R_{v}^3}{\kappa}, \quad \quad \text{for vesicles} \\
%&\tau_c =\frac{\mu R_{v}}{\zeta_0}, \quad \quad \text{for drops}.
\label{relaxation}
 \end{equation}
For vesicles, the Lagrange multiplier is non-dimensionalized  as
\begin{equation}
\zeta^* = \zeta R_{v}^2/\kappa.
\end{equation}
%The radius $R_{v}=\sqrt{S/\pi}$ (with $S$ the enclosed area) of the particle is taken as  characteristic length scale. %
The dimensionless integral equation of the velocity along the contour is given by
\begin{equation}
C_a \mathbf{u}^*(\boldsymbol{x}^{*}) = C_a \mathbf{u}^{\infty *}(\boldsymbol{x}^{*}) + 
\frac{1}{4\pi} \sum_{i}
\int\limits_{{\partial \Omega_{i}}}{\mathbf{\mathcal{G}}}^{2W}(\boldsymbol{x}^{*},\boldsymbol{y}^{*}) \cdot \mathbf{f}^{*}(\boldsymbol{y}^{*})  ds^*(\boldsymbol{y}^{*})
\label{eq::interfacialflowfinal}
\end{equation}
 The dimensionless expression $\mathbf f^*$ of the membrane force reads
\begin{equation}
\mathbf{f}^{*} =
\begin{cases}
   \left(\frac{\partial^2{c^*}}{\partial {s^*}^2} + \frac{{c^*}^3}{2}\right)\mathbf{n}  - c^* \zeta^* \mathbf{n} +
\frac{\partial{\zeta}^*}{\partial{s^*}}\mathbf{t} \quad & \text{for vesicles} \\
- c^* \mathbf{n} \quad & \text{for drops.}

\end{cases}
\label{eq::membrane_force_dimensionless}
\end{equation}
In addition to the reduced area defined above for vesicles ($\nu =(S/\pi)/(p/2\pi)^2$), %and the confinement given by $C_n=\frac{2R_v}{W}$, 
we have a new dimensionless number associated to the flow in equation (\ref{eq::interfacialflowfinal}), which is given by
\begin{equation}
C_a = 
\begin{cases}
\frac{\mu R_{v}^3\dot{\gamma}}{\kappa}=\frac{\mu R_{v}^4}{\kappa}\frac{u_{max}}{(W/2)^2}\quad & \text{for vesicles}\\
\frac{\mu R_{v}\dot{\gamma}}{\zeta_0}=\frac{\mu R_{v}^2}{\zeta_0}\frac{u_{max}}{(W/2)^2}\quad & \text{for drops.}
\end{cases}
\label{capi}
\end{equation}
We refer to this number as  the capillary number. Here $\dot{\gamma}=4 {R_v}u_{max}/{W^2}$ is the shear rate defined as the value of $\partial u_1^{\infty} /\partial x_2$ at  $x_2=R_v/2$. We recall that the shear rate in a parabolic flow is position-dependent, unlike in a linear shear flow. Thus, the definition of the capillary number for a parabolic flow is not unique
 in the literature. For example, in \cite{kaoui2011complexity} the chosen shear rate is that at the wall and is equal to $\dot\gamma_W=4u_{max}/W$. The ratio between the present  shear rate  and that in \cite{kaoui2011complexity} is equal to $R_v/W$. In a previous study \cite{Othmane2014} as a validation of the present code, we have reproduced the full phase diagram of the vesicle shapes obtained in Ref.\cite{kaoui2011complexity} in the plane of capillary number and the degree of confinement. Here, we shall explore the ranges of parameters where the vesicle shape is of parachute type only (slipper shapes are excluded from our study). Similarly, the drops are maintained in the center of the channel in order to make an adequate comparison with the vesicle case.
%Note that for drops the bending mode  is absent and we have to introduce a capillary number based on surface energy (see section \ref{sec:drops}).

The capillary number is the ratio between the flow stress and bending force density. It may be viewed also as the ratio between the characteristic shape relaxation time $\tau_c$ (eq. (\ref{relaxation})) and the time scale of the flow $\tau_f = 1/\dot{\gamma}$. In order to have a reference for the conversion of dimensionless units into physical ones, the following dimensional numbers for RBCs can be used: $R_v=3\mu m$, $\mu=10^{-3}$ Pa$\cdot$s and $\kappa=10^{-19}J$. This leads to a characteristic time of shape relaxation of about $\tau_c \sim 0.2-0.3 $ s. This is quite consistent with measured values for RBCs\cite{hochmuth1979red,tomaiuolo2011start,prado2015viscoelastic}.

{In this study, we quantify the hydrodynamic interactions of two particles by tracking the time evolution of the distance between their centers of mass. Let us denote the
leading particle by $1$ and the following particle by $2,$ and let $\mathbf X^{(i)}(t)=R_v\mathbf X^{*(i)}(t)$ denote the instantaneous position of the center of mass of the contour of particle $i$ at time $t.$ Then the distance between the particles (called interdistance below) is defined as \begin{equation}\label{eq:deltaX}\Delta X(t)=R_v \Delta X(t)^*=X_1^{(1)}(t)-X_1^{(2)}(t).\end{equation} Consistently, we define the velocity with which the particles approach (or separate from) each other as \begin{equation}\label{eq:deltaU}\Delta U(t)=\frac{R_v}{\tau_c}\Delta U(t)^*=\frac{d\Delta X(t)}{dt}.\end{equation}
}

\subsubsection{Numerical method for the vesicle dynamics}
The integral equation (\ref{eq::interfacialflowfinal}) is discretized using the trapezoid rule and derivatives are approximated using a finite difference scheme. Each particle is described by a collection of equispaced Lagrangian nodes advected by the flow. Their motion is obtained by solving the advection equation for each material node $\boldsymbol{x}$ lying on the membrane
\begin{equation}
\frac{d\boldsymbol{x}}{dt} = \mathbf{u}(\boldsymbol{x})
\label{eq:material}
\end{equation}
We calculate the velocity of each discretized point on the membranes by solving the integral equation (\ref{eq::interfacialflowfinal}), and the position of each node is updated at each time step using an explicit Euler scheme.
\begin{equation}
\boldsymbol{x}(t+dt) = \boldsymbol{x}(t) + \mathbf{u}(\boldsymbol{x}(t),t) dt
\end{equation}
A tension-like parameter is introduced as a penalty parameter instead of the Lagrange multiplier $\zeta$ (which enters both tangential and normal force; see equation (\ref{eq::membrane_force})). In other words, each material point is linked to each of its two neighbors via a very stiff "spring" in order to enforce the local conservation of arclength, as described in \cite{ghigliotti2010rheology}.  If the spring stiffness is denoted as $T_{tens}$, we can define a time scale $\tau_{tens}=\mu/(T_{tens} R_v)$. This numerical time scale is to be compared to the physical time scale $\tau_c$ defined in equation (\ref{relaxation}) and to the flow time scale $\tau_f=\dot{\gamma}^{-1}$. $\tau_{tens}$ must be taken small enough in comparison to $\tau_c$ and $\tau_f$ so that on the physical and flow time scales the local incompressibility of the membrane is safely satisfied. For most practical purposes $\tau_{tens}=10^{-4}-10^{-3} \tau_c$ has proven to be largely sufficient (see below). %

There is no arclength conservation for drops. Therefore, the discretization points tend to accumulate in some regions of the membrane while depleting in the others if the simple advection of material points (\ref{eq:material}) is used. We resolve this challenge by applying an additional displacement of the discretization points every time step. The displacement field is chosen to be (i) tangential to the interface of the drop so that its shape remains the same and (ii) such that the discretization points approach the equispaced distribution of the interface. 

Each membrane is described by $N_{mem}=120$ nodes whose positions are updated each $\Delta t = 10^{-4}\tau_c$ (i.e. this corresponds to the time step). The relative errors corresponding to  the area, the perimeter and the reduced area are around $0.07\%$, $0.035\%$, and $0.0009\%$, respectively. The steady-state value of the distance between the mass centers of the two vesicles $\Delta X^*_f\equiv\Delta X^*(\infty)$ is also reported (see Table.\ref{tab:accuracy}). The calculations are performed on a cluster consisting of 32 dual-core AMD64 processors with 24GB RAM per node. OpenMP directives are used to parallelize the matrix-vector product computation. The time needed to complete $10^6$ iterations as a function of the number of cores is reported in Fig.\ref{scalability_cpus}(a) using two vesicles in a channel of width $W/R_v=2$ and a $C_a=10$. Similarly, we have plotted the required run time to complete $2\cdot10^{5}$ steps using 12 cores as a function of the number of vesicles (see Fig.\ref{scalability_cpus}(b)). It is important to underline that some of the cases reported in the phase diagram in the result section ran over more than two weeks on a 12-core node since we decided to avoid using any cut-off or periodic boundary conditions in our system due to the long-range nature of the hydrodynamic interaction. The use of the appropriate Green's function (that vanishes on the walls) allowed us to avoid finite size  effects, since we can consider  literally  an  infinite domain along the flow direction.

\subsubsection{Numerical method for the velocity field inside and outside the particles}
The velocity field in the fluid domains (inside and outside the particles) obeys the same boundary integral equation as that on the membrane (Eq. (\ref{eq::boundary_integral})), where now $\boldsymbol{x}$ is a location of any point in the $(x_1,x_2)$ plane. Once a steady-state configuration is reached (the final shape as well as the vesicle interdistance), the velocity field is
evaluated as a post-processing task. We introduce a regular square grid (with a certain degree of refinement; the mesh size can be taken significantly smaller than $ds$ if need be). Since the Green's function is singular when the target point coincides with the source point, a small stripe (of order $ds$ in width) around the membranes is excluded from the fluid domain in order to ensure a good behavior of the velocity field. The lattice points are in general not on the membrane, and we only need to evaluate the distance between the source point (lying on the membrane) and the target point (lying on the square grid). For each point $\boldsymbol{x}$, the velocity field is evaluated by using Eq. (\ref{eq::boundary_integral}), where the integral along the membrane is performed exactly in the same way as in the previous section.

%
%\begingroup
%\squeezetable
\begin{table}
\caption{\label{tab:accuracy}{Relative errors on area, perimeter and reduced area of vesicles in Poisuille flow $N_{mem}=\{120; 180; 240\}$ and $\Delta t = 10^{-4} \tau_c.$ Also shown is the final distance between the mass centers of the vesicles $\Delta X^*_f.$ The other relevant parameters are $C_a=10$ and $W=2R_v$. We obtained almost the same results for $\Delta t = 5.10^{-5} \tau_c$.}}
\begin{ruledtabular}
\begin{tabular}{lllll}
 $N_{mem}$& ${\frac{(S-S_0)}{S_0}} (\%)$ & ${\frac{(p-p_0)}{p_0}} (\%)$ & ${\frac{(\nu-\nu_0)}{\nu_0}} (\%)$ & ${\Delta X^*_f}$\\
\itshape ${120}$   & $0.07$    & $0.035$ & $0.0009$ & $5.83$\\
\itshape ${180}$   & $0.03$    & $0.015$ & $0.0004$ & $5.83$\\
\itshape ${240}$   & $0.017$   & $0.008$ & $0.0003$ & $5.83$\\
\end{tabular}
\end{ruledtabular}
\end{table}
%\endgroup
%
 \begin{figure}[h!]
        \centering
        \includegraphics[width = 1. \linewidth]{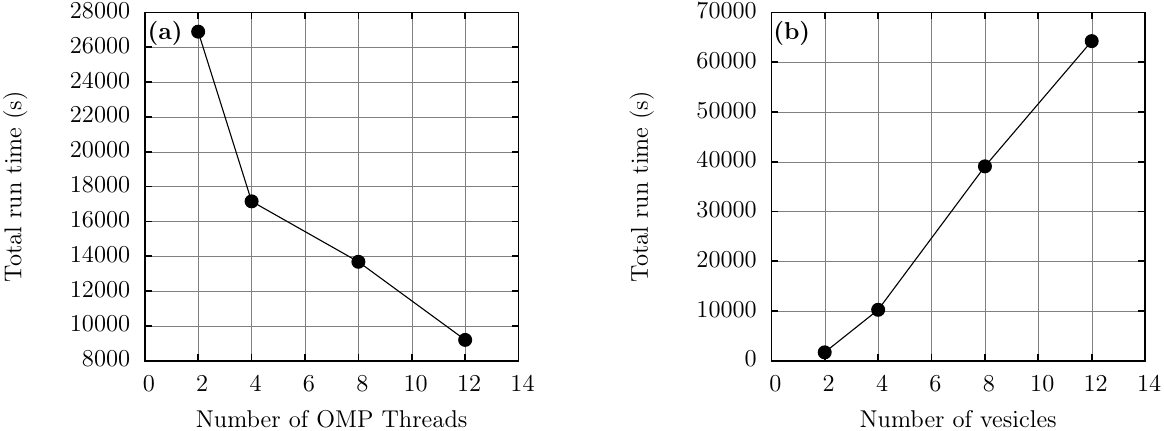}
        \caption{{The Total run time (in seconds) required to compute $10^6$ iterations as a function of the number of cores (a). The total run time needed to reach $2\cdot10^{5}$ steps as a function of the number of vesicles using 12 cores (b).}}
        \label{scalability_cpus}
 \end{figure}
}
\section{Results for vesicles}
\label{sec:results}
In this section, simulations of a pair of vesicles in channels of different widths are performed. Unless indicated otherwise, vesicles are initialized as elongated ellipses with major axis parallel to the flow direction. {We quantify the hydrodynamic interaction between the vesicles by tracking the interdistance $\Delta X^*$ } %
We investigate the role of {the confinement,} the capillary number, and the initial configuration (shapes and interdistances of vesicles) on the final state. {We shall see that in some cases, there are several coexisting stationary interdistances for a given value of confinement and capillary number.}
 \subsection{Weak confinement}
 A previous study has been devoted to the cluster  formation in the absence of  walls \cite{Giovanni2012}. %
 We first study the behavior of a pair of vesicles in weak confinement in order to check that we can capture almost the same result. %
 We have analyzed the time evolution of a pair of vesicles in a channel having a width $W=20R_v$, which corresponds to weak confinement. %
 We find that the steady-state interdistance is equal to about $2.53R_v$ for $W=20R_v$. This result compares well with that obtained in an unbounded flow \cite{Giovanni2012} where the steady-state interdistance is of about $2.4R_v$.

 We have analyzed systematically the behavior of a pair of vesicles for different {(but still weak)} confinements and different initial conditions. %
 Figure \ref{13p1} shows a typical behavior of the pair interdistance as a function of time for different initial separations, which are denoted as $\Delta X^*_{init}\equiv\Delta X^*(0)$ in that figure. %
 We see there that different initial conditions lead to the same final state.
Figure \ref{floww20} shows the final configuration of the pair of vesicles as well as the induced flow field, that is the total flow field from which we subtract the imposed Poiseuille flow. The total flow field is also shown.

 \begin{figure}[h!]
 	\centering
 	\includegraphics[width = .65 \linewidth]{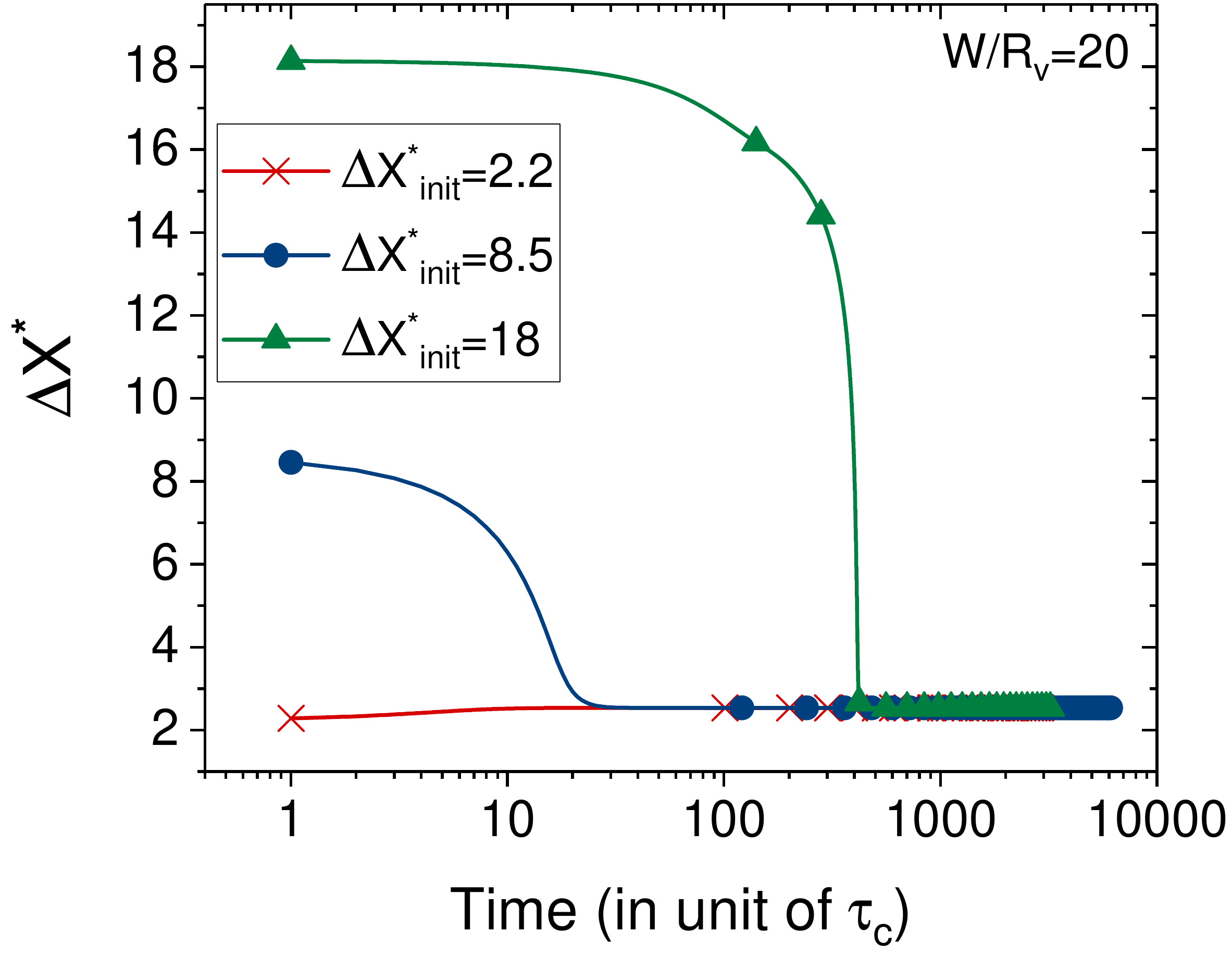} 	
 	\caption{The pair interdistance as a function of time for different initial conditions (initial interdistance). $W=20R_v$, and $C_a=10$. %
 	Note that the horizontal axis is represented in log scale.}
 	\label{13p1}
 \end{figure}

 \begin{figure}
 	\centering
 	\includegraphics[width = 1. \linewidth]{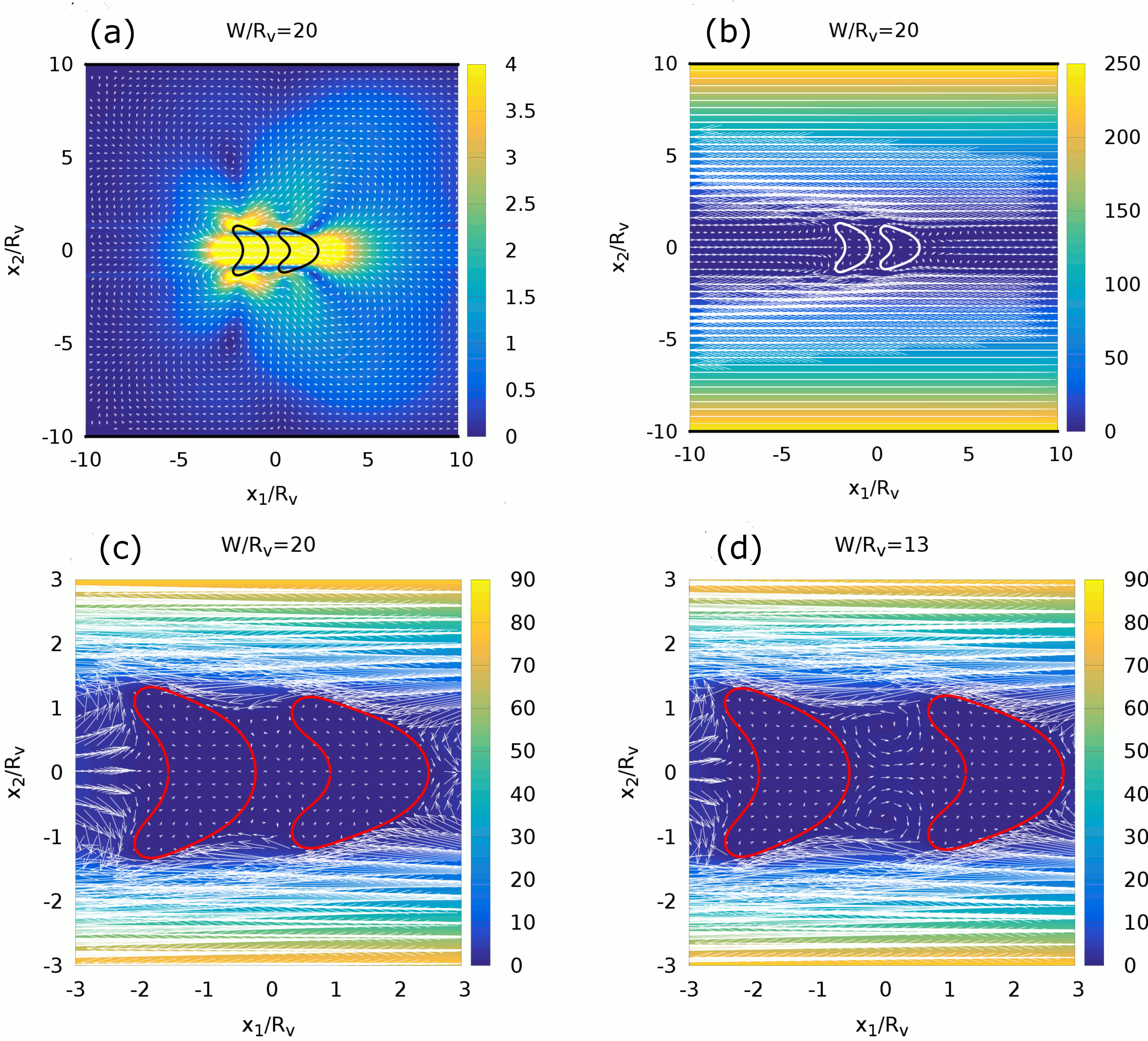}
 	\caption{ The induced flow field for a pair of vesicles in a weakly confined flow (a); $W=20R_v$. %
 	We also represent the flow in the frame moving with the vesicle called hereafter \textquotedblleft co-moving frame \textquotedblright (b). %
 	A zoom in  the co-moving frame in the region located between the cells reveals the absence of bolus in case of $W=20R_v$ where the final interdistance is about $2.4R_v$ (c); %
 	and the presence of quasi-circular bolus for $W=13R_v,$ where the  final interdistance is of about $3.4R_v$ (d).}
 	\label{floww20}
\end{figure}

In order to further analyze the time evolution of vesicle pairs, we have studied systematically the behavior of the relative velocity of the two centers of mass of the pair of vesicles $\Delta U^*$ as a function of their interdistance $\Delta X^*$ by considering different initial interdistances. 
 A positive value of $\Delta U^*$ means that the two vesicles repel each other, while a negative value means they attract each other. A stationary interdistance corresponds to $\Delta U^*=0$.
 \begin{figure}[h!]
 	\centering
 	\includegraphics[width = .65 \linewidth]{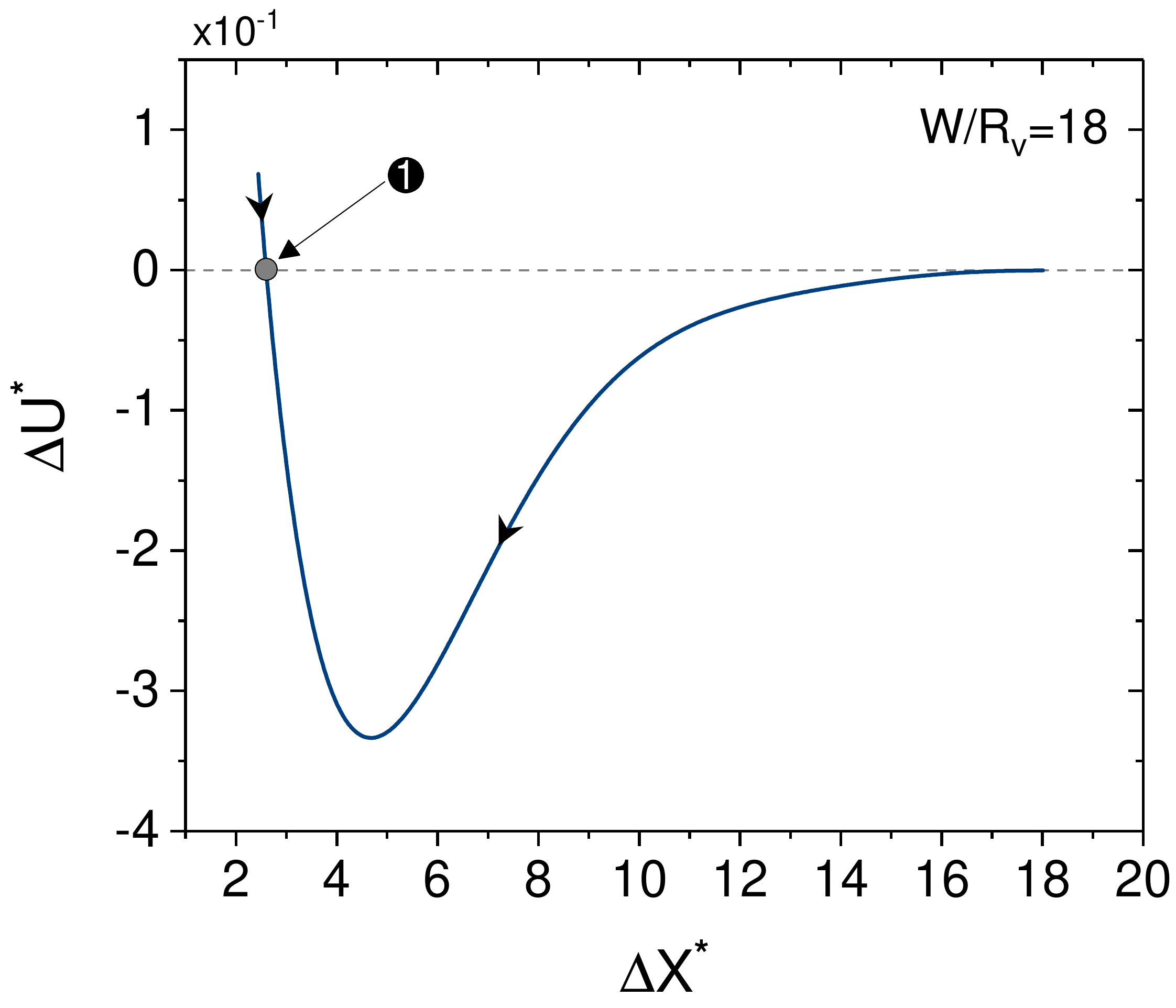}	
 	\caption{ The relative velocity as a function of the interdistance; $W=18R_v$, and $C_a=10$.}
 	\label{vander1}
\end{figure}
 Figure \ref{vander1} shows ${\Delta} U^{*}$ as a function of the dimensionless interdistance between centers of mass, ${\Delta} X^{*}$.
 We observe a repulsion at short interdistance and an attraction at long interdistance. {There exists one stationary interdistance characterized by ${\Delta} U^{*}=0$ (denoted as $1$ within a dark circle in Figure \ref{vander1}). This stationary solution is unique and independent of the choice of the initial shapes explored so far (Figure \ref{vander12}).
 \begin{figure}[h!]
        \centering
        \includegraphics[width = 1. \linewidth]{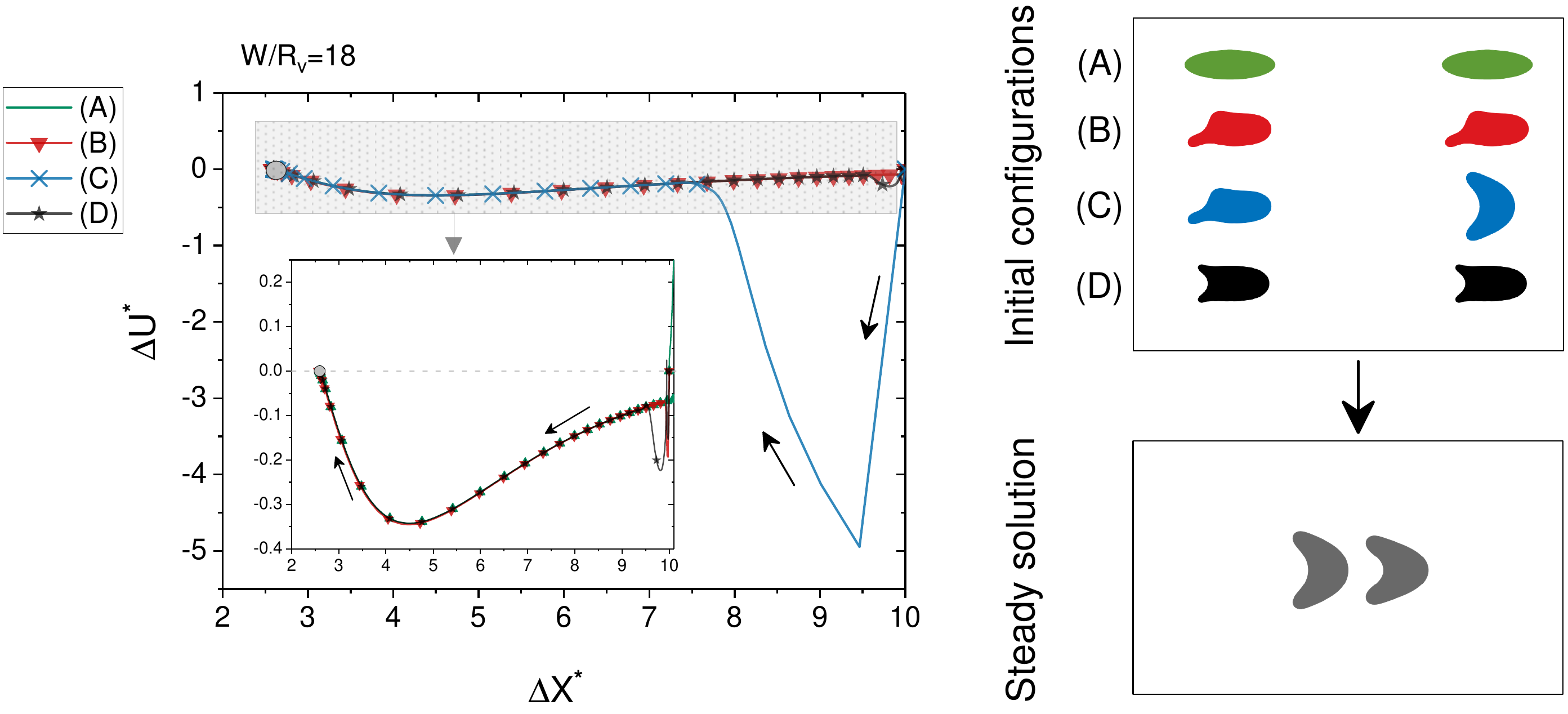}
        \caption{{The relative velocity as a function of the interdistance; $W=18R_v$, and $C_a=10$. Starting from different initial shapes, the pair of vesicles converges to the same stationary solution. The labels (A), (B), (C), and (D) correspond to the initial shapes depicted in the right figure.}}
        \label{vander12}
\end{figure}
Furthermore, this stationary solution is unambiguously stable since for longer interdistances the velocity is negative meaning the vesicles attract each other, while for shorter interdistances, the relative velocity is positive and the vesicles repel. Generally, if the relative velocity crosses zero by going from positive to negative values (as $\Delta X^*$ increases), we have a stable stationary solution. 

We performed this study for several other weak confinements and determined the corresponding stationary interdistances.} %
\begin{figure}[h!]
 	\centering
 	\includegraphics[width = .5 \linewidth]{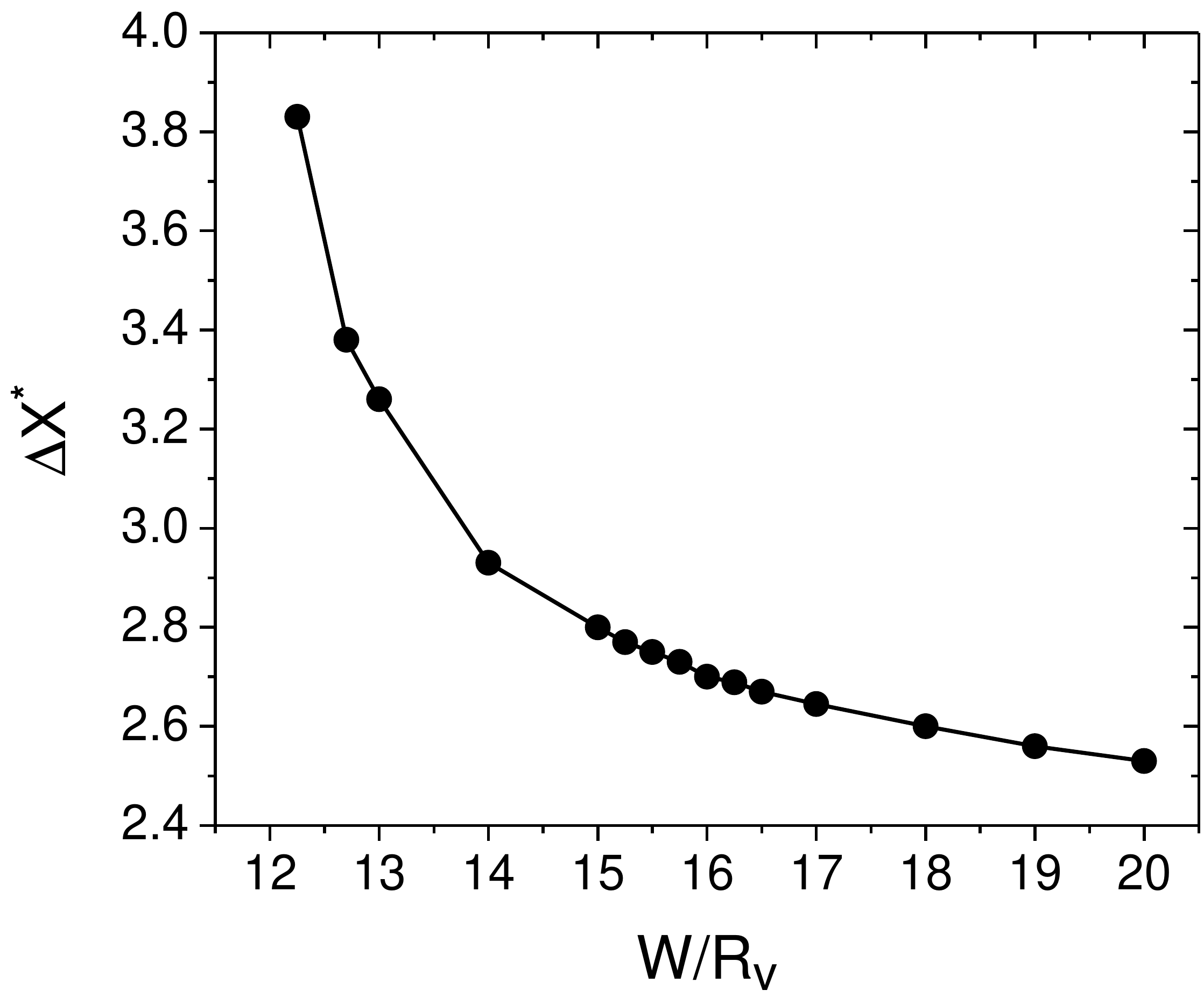}	
 	\caption{The branch of solution representing the stationary interdistance as a function of the channel width. }
 	\label{diag1}
\end{figure}
Figure \ref{diag1} shows the branch of stationary interdistance ${\Delta} X^{*}$ as a function of $W/R_v$.  The stationary interdistance weakly depends on confinement, and remains close to about $3R_v$.

  \subsection{Strong confinement}
{ Let us now examine the generic behavior under strong confinement. Consider the case $W=3R_v$. Looking at the evolution of the interdistance as a function of time,  we find that the vesicle pair settles into a bound, stable steady state for several initial conditions. Two examples are illustrated in Fig. \ref{W31}. The first noticeable feature is a significantly larger stationary interdistance for $W=3R_v$ than for weak confinement: the stationary interdistance of about $5R_v$ for $W=3R_v$ is about two times the value found for weak confinement (about $2.5R_v$). %
  A first interpretation would suggest the screening of the hydrodynamical interaction by the confining walls.
However, since the screening is felt both in attraction and repulsion, this reasoning is  {\it a priori} not justified, leaving unclear the mechanism by which confinement would shift both attraction and repulsion zones.} %

  \begin{figure}[h!]
 	\centering
 	\includegraphics[width = .95 \linewidth]{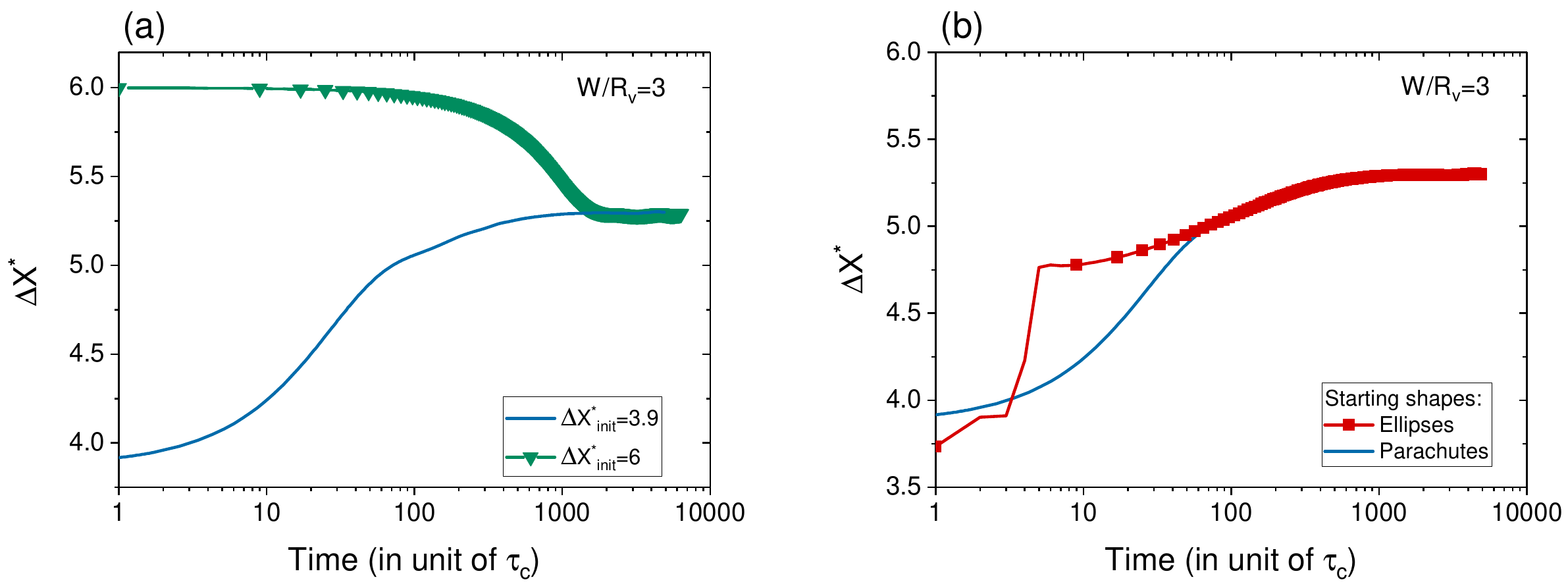}	
 	\caption{{The interdistance of a pair of parachute-like vesicles as a function of time for different initial interdistances $\Delta X^*_{init} = 3.9$ (solid line) and $\Delta X^*_{init} = 6$ (symbols). (b): The effect of the initial shape of vesicles in the pair on the steady-state solution is shown for ellipses (symbols) and parachutes (solid line). $W=3R_v$, $C_a=10$. Note that the horizontal axis is represented in log scale.}}
 	\label{W31}
  \end{figure}

To approach a more rigorous interpretation, let us first analyze the velocity field in a strong enough confinement regime. Figure \ref{vW63} shows the induced flow field for two confinements $W=6R_v$ and $W=3R_v$, as well as the flow field in the co-moving frame (the frame moving with the pair). %
   \begin{figure}
 	\centering
 	\includegraphics[width = 1. \linewidth]{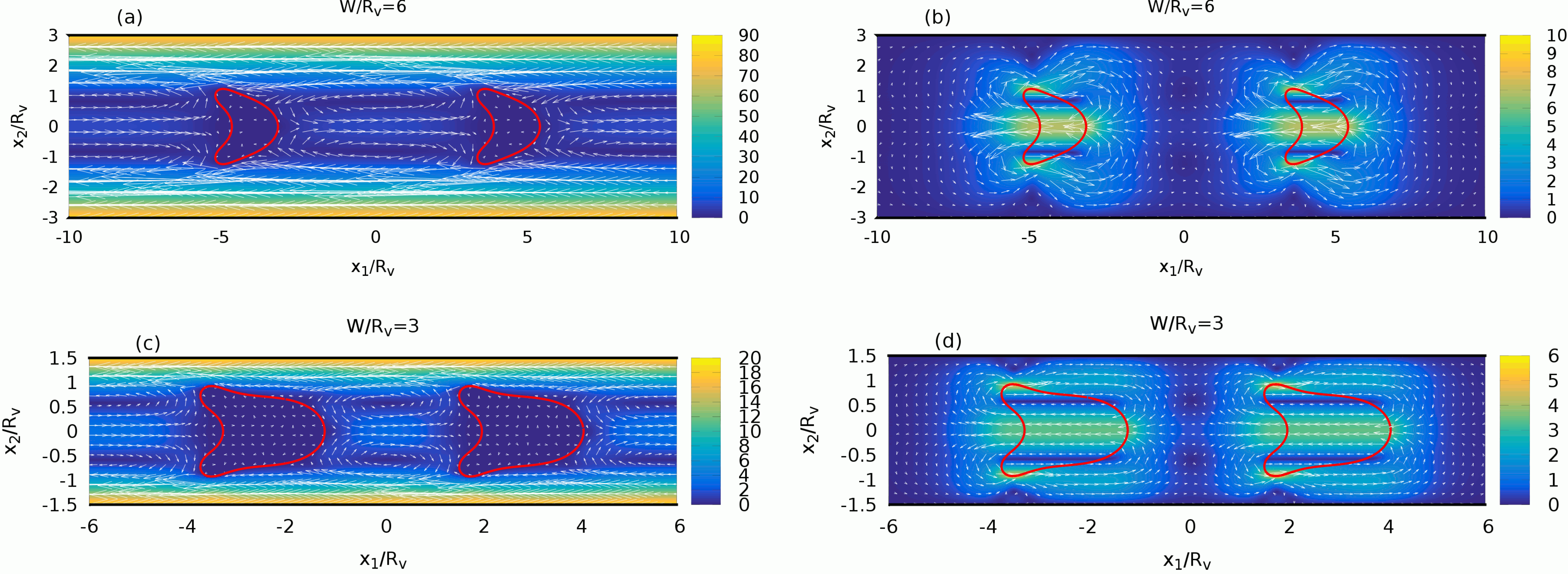}
 	\caption{ The flow field in a co-moving frame (a,c) and the induced flow (b,d) field for a pair of vesicles for $W=6R_v$ and $W=3R_v$.}
 	\label{vW63}
   \end{figure}
 The main difference is that the flow lines which extend far away in the case of weak confinement (see Figure \ref{floww20}) are cut off by the effect of the walls in the case of strong confinement. As we shall see below, this partially hints to a weaker interaction magnitude. %
As for the weak-confinement case, we also analyze the relative velocity as a function of the interdistance. The result is shown in Figure \ref{vander2}.
\begin{figure}[h!]
 	\centering
 	\includegraphics[width = .5 \linewidth]{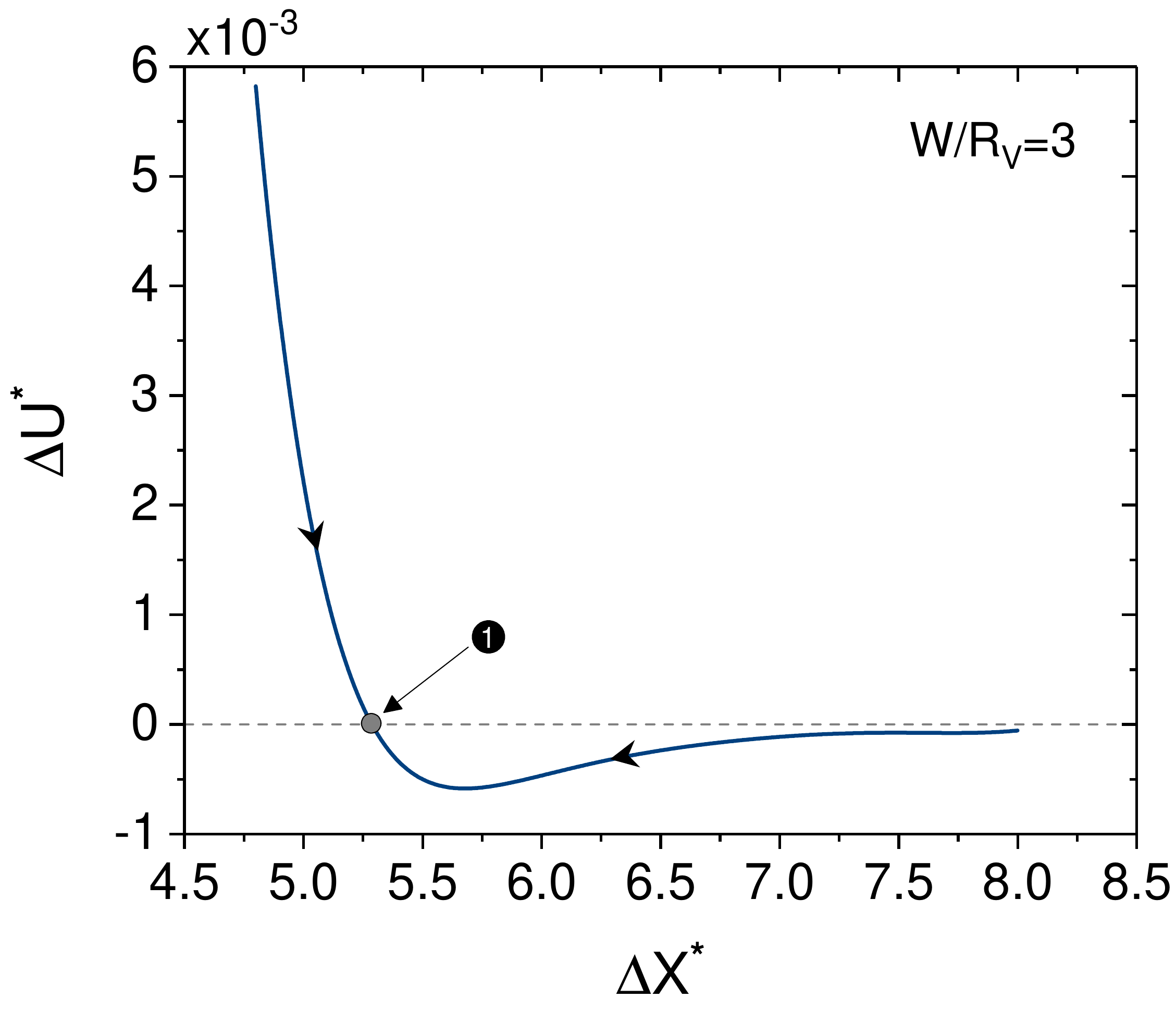}	
 	\caption{ The relative velocity as a function of the interdistance; $W=3R_v$, and $C_a=10$.}
 	\label{vander2}
\end{figure}
  We obtain a stable stationary solution, which results for a short-range repulsion and long-range attraction.
  We find that the relative velocity amplitude is  significantly smaller than  that obtained for a weak confinement (compare with Figure \ref{vander1}). { This can be traced back to the screening effect of the walls which weaken the interaction. We perform this study for several confinements to determine the corresponding steady-state solution. }
\begin{figure}[h!]
 	\centering
 	\includegraphics[width = .5 \linewidth]{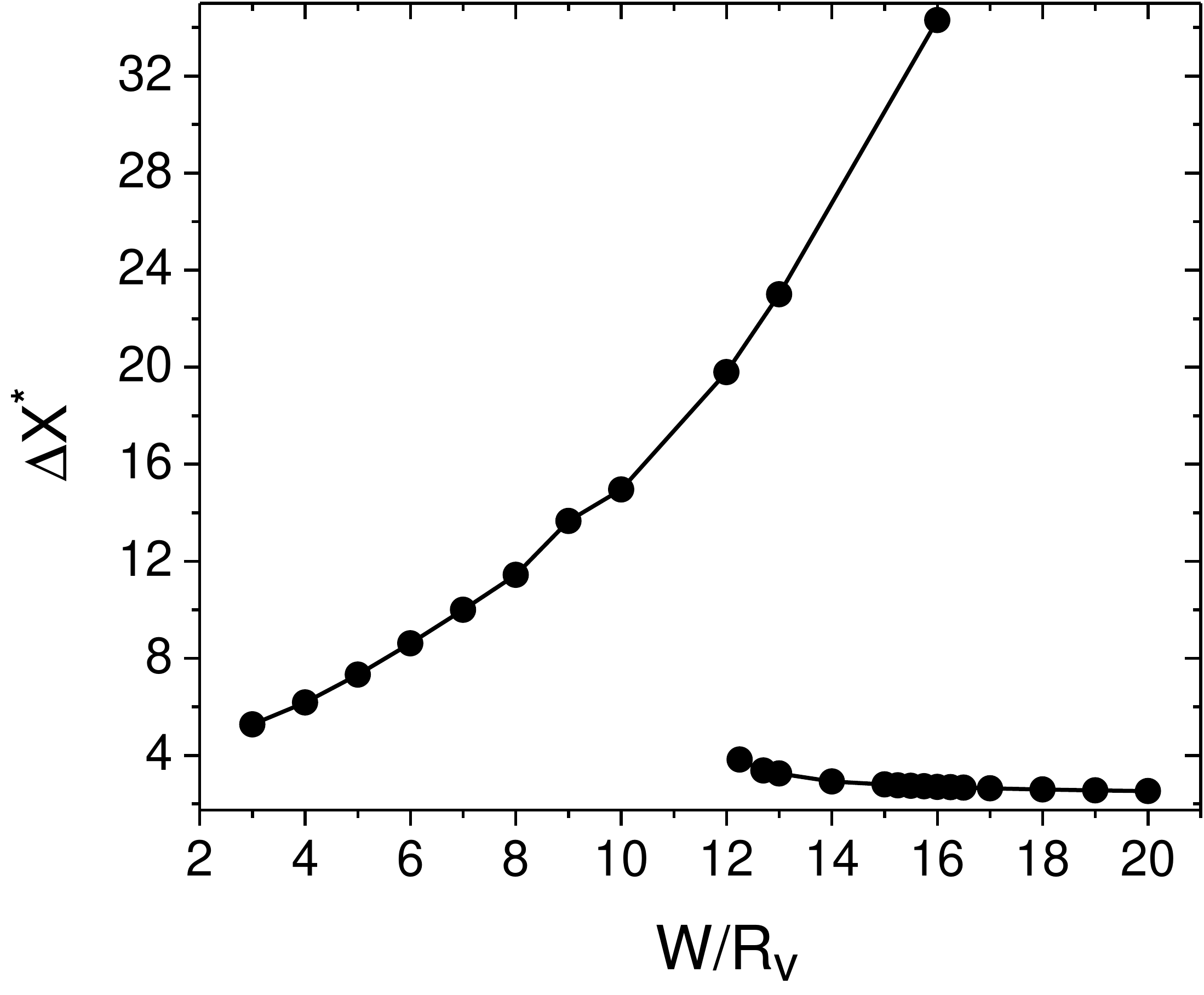}	
 	\caption{The branch of solution representing the stationary interdistance as a function of the channel width for a strong confinement.}
 	\label{diag2}
\end{figure}
Figure \ref{diag2} shows the branch of stationary interdistance $\overline{\Delta X}$ as a function of $W/R_v$.
In contrast to the weakly confined case (also shown in Figure \ref{diag2}), the stationary interdistance strongly depends on confinement, and varies from about $4R_v$ up to about $32R_v$. { We have a new solution branch for small values of $W$ (the leftmost branch in Figure \ref{diag2}), distinct from the one discussed above for large $W$ (the rightmost branch in Figure \ref{diag2}). We have two distinct  branches, one presenting a stationary interdistance that increases with $W$ and the other whose stationary interdistance decreases with $W$. We shall now dig further into the structure of the phase diagram.} %
 \subsection{Full phase diagram}
{ We broaden our investigation in order to clarify the overall structure of the topology of the phase diagram. First, we  analyze the basin of attraction of each branch by exploring a wider range of initial conditions. %
 We begin our discussion with the weak-confinement case. In the previous section, we saw that different initial conditions led to the same final solution. Exploring wider and wider regions of initial conditions reveals a different scenario. Keeping the same confinement $W=13R_v$, as above, we find that beyond a certain initial pair interdistance, the solution no longer converges to the same value.} %
   \begin{figure}[h!]
 	\centering
 	 \includegraphics[width = .8 \linewidth]{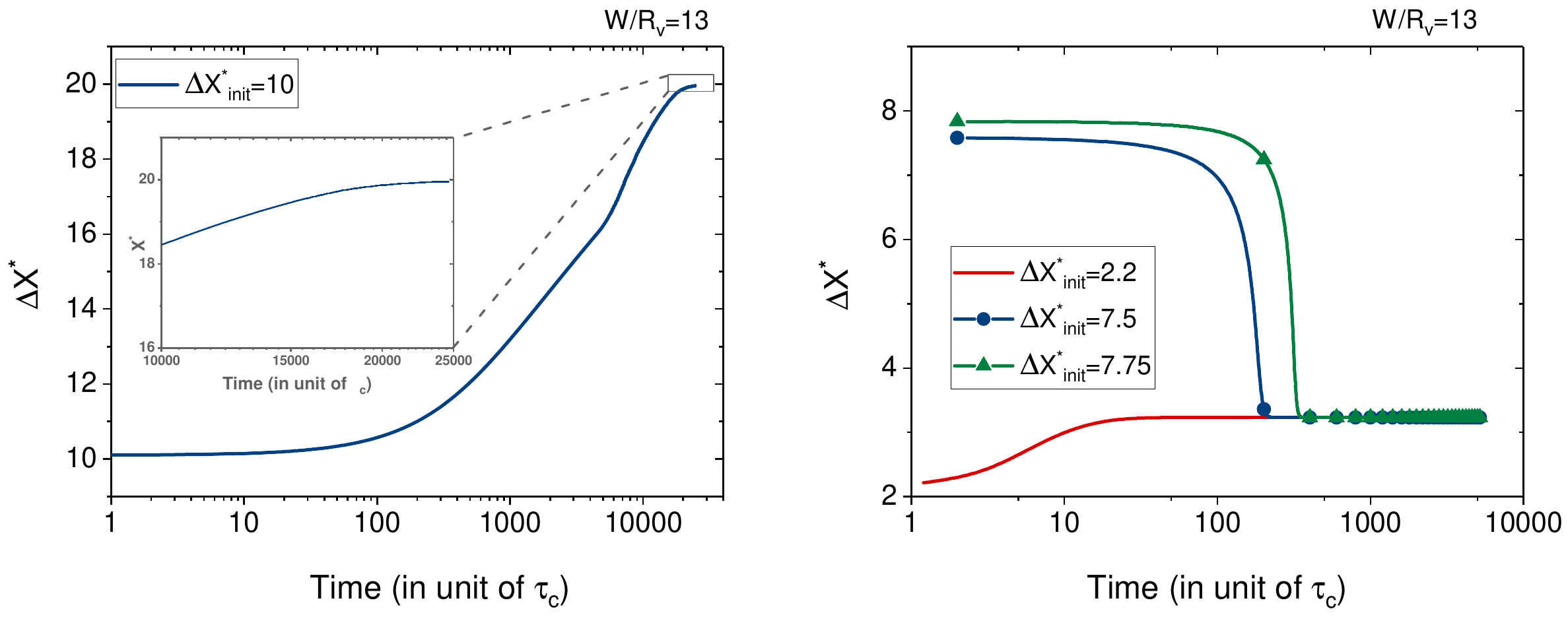}
 	\caption{ The pair interdistance as a function of time for different initial interdistance: $\Delta X^*_{init} = 10$ (left); and $2.2,$ $7.5,$ and $7.75$ (right). $C_a=10$ and $W/R_v=13$. %
 	Note that the horizontal axis is represented in log scale.}
 	\label{13p2}
\end{figure}
 Figure \ref{13p2}(left) shows the {time evolution of the interdistance}, which is still evolving after $25000\tau_c$ but converges ultimately to a value of about $23 R_v$. %
{ For different initial conditions (shorter initial interdistances), we have seen a final interdistance of about $3.4 R_v$ (Figure \ref{13p2}(right)). This clearly demonstrates the coexistence of two different stable solutions. In general, two stable solutions should be separated by an unstable solution. We seek to determine the location of the unstable branch by analyzing the relative velocity as a function of interdistance, as before.} %

 \begin{figure}
 	\centering
 	\includegraphics[width = .75 \linewidth]{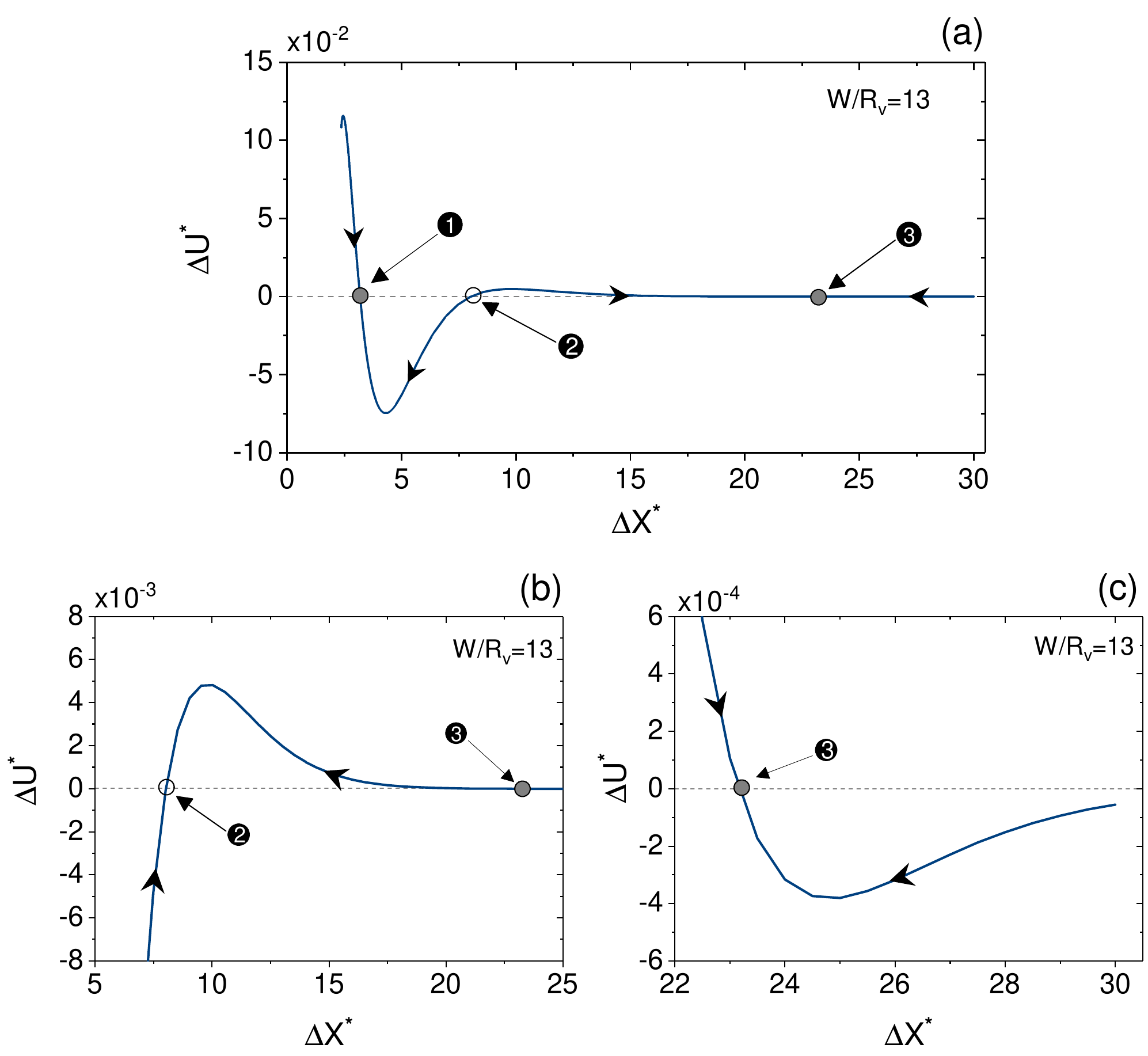}
 	\caption{ Normalized relative velocity as a function of the interdistance for $W/R_v=13$. (a), (b) and (c) show successive zooms.  Solid dots and open circles represent, respectively, stable and unstable fixed points. %
 	A succession of stable and unstable fixed points is observed.}
 	\label{vandercom}
\end{figure}

{ Our results are shown in Figure \ref{vandercom}. The locations where the relative velocity of the vesicles vanishes are indicated by points marked 1, 2, and 3. Points 1 and 3 correspond to a stable interdistance, whereas point 2 corresponds to an unstable one. Since the relative velocity in the vicinity of point 3 is very small, we zoom in to reveal the structure of the dependence of the relative velocity on the interdistance. Another systematic analysis done by varying the confinement (i.e. $W$) allows us to show the full diagram of stationary solutions (like points 1, 2, and 3) as a function of $W$. }
The results are summarized in Figure \ref{diag_full}. %
 \begin{figure}[h!]
 	\centering
 	\includegraphics[width = .65 \linewidth]{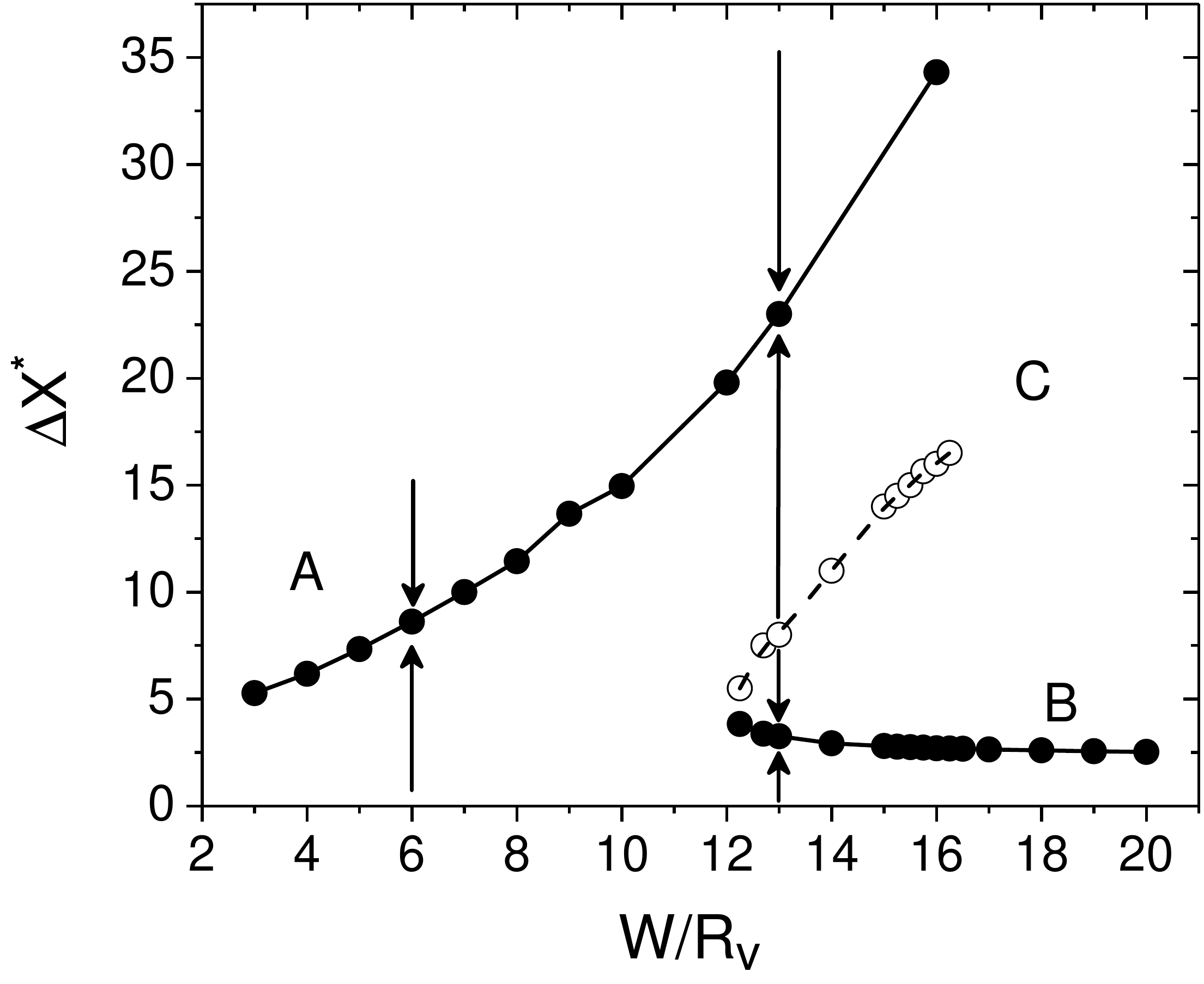}	
 	\caption{The full set of branches of stationary interdistance as a function of the channel width. Solid lines represent stable branches, whereas dashed lines unstable ones.}
 	\label{diag_full}
\end{figure}
{ We see that the branch {for weak confinement} undergoes a fold singularity in the form of a saddle-node bifurcation, in which a stable solution (represented by solid line) merges with an unstable solution (dashed line). The branch arising at {strong confinement} continues to exist (as a stable solution) beyond the saddle-node point, and does not show, for the values of $W$ explored so far, any sign of disappearance. Theoretically, this branch may either continue to exist for any $W$, or it may undergo a fold singularity, one of the scenarios expected from catastrophe theory. Since for {weak confinement} the relative velocity becomes too small { to be of practical interest}, the question of {the behavior of the branches at long interdistances} is only academic. We shall thus not dwell here any farther on this issue. %

Finally, let us say few words on the strength of interaction as a function of confinement. As can be expected, confinement reduces the strength of the interaction by screening. To understand the difference in the pairing mechanism at large and small channel widths we first consider a pair of vesicles in two different channels of widths $6R_v$ and $15.25 R_v$. The time evolution of the interdistance in each channel is shown in Figure \ref{fig::Fig16}.} %
\begin{figure}[h!]
 	\centering
 	\includegraphics[width = .95 \linewidth]{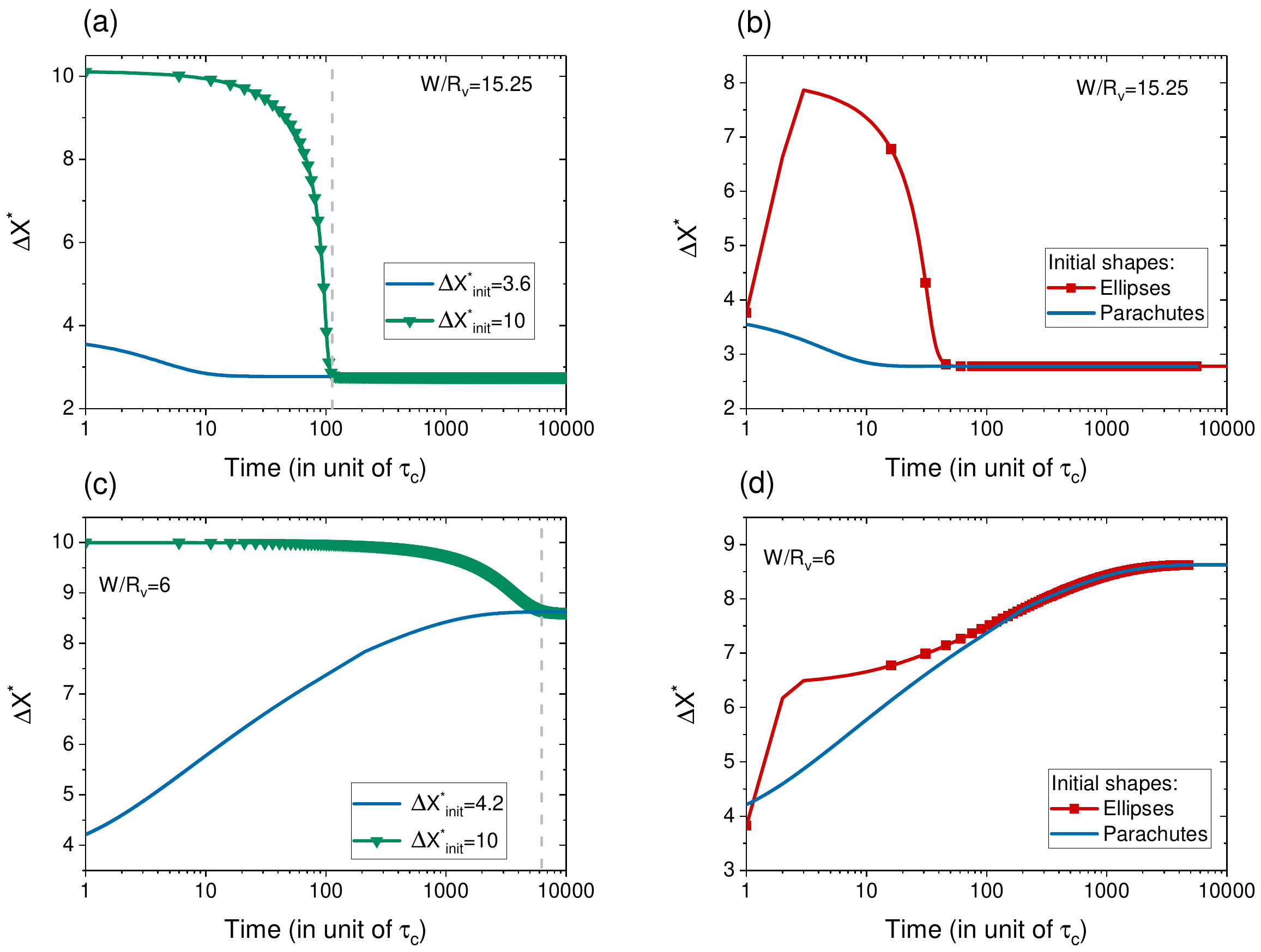} 	
 	\caption{{Time evolution of the interdistance between two vesicles in different channel widths. (a) and (b) $W=15.25R_v$. (c) and (d) $W=6R_v$. In (b) and (d), different initial shapes ellipses (symbols) and parachutes (solid line) are used. Note that the horizontal axis is represented in log scale.}}
        \label{fig::Fig16}
\end{figure}
We looked at characteristic time needed to reach the steady state for different channel widths finding that, for the channel of width $W=6R_v$, the interdistance reduces by approximatively $1.4R_v$ in a time $7000\tau_c,$ whereas in a channel of width $W=15.25R_v$, the interdistance reduces by $7.2R_v$ in only $120\tau_c$. Considering that the typical time $\tau_c$ for RBC is about $0.1$ s gives a time of about 10 min in the first case and 10s in the second one}. We must note that in reality there are always fluctuations and imperfections keeping the cells from remaining in the same lateral position. For example, if the leading cell is slightly off-centered its velocity may be sensibly different from that of the following cell, so that the characteristic time needed to reach the bound state may be on a significantly shorter time scale than $10$ s. %
%%%
\subsection{Effect of the capillary number} \label{ssec::res_cap}
\begin{figure}[h!]
 	\centering
 	\includegraphics[width = 1. \linewidth]{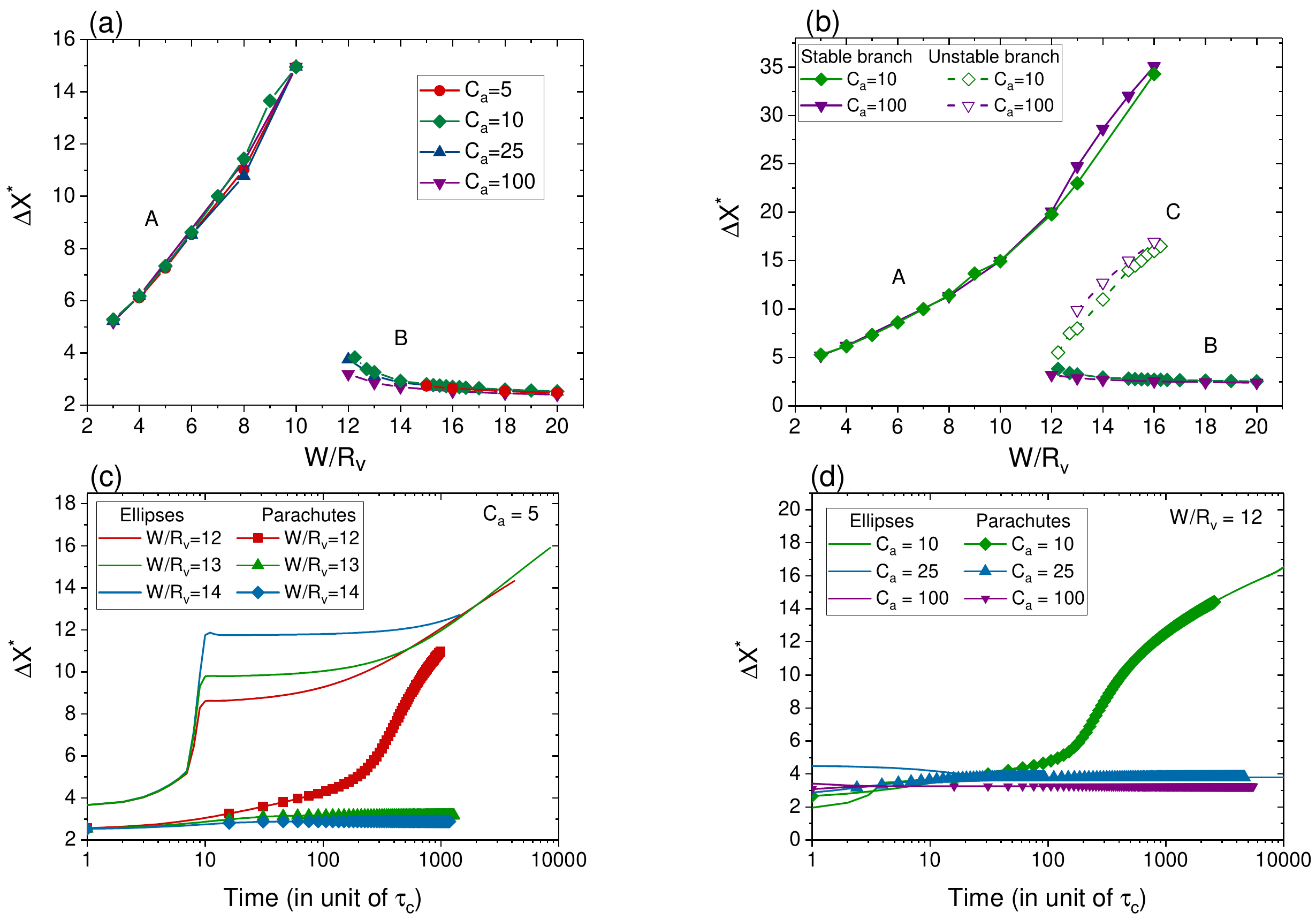}	
 	\caption{{Stationary interdistance as a function of the channel width for different capillary numbers (a). The phase diagram for $C_a = 10$ and $C_a = 100$ (b). Role of the initial shape (solid lines are ellipses and dashed lines are parachutes) on the time evolution of the pair interdistance for $C_a=5$ and $W/R_v = 12$, $13$ and $14$ corresponding to the transition area (c). Same as (c) but for $W/R_v =12$ and $C_a=10$, $25$, and $100$ (d). Note that the horizontal axis is represented in log scale in panels (c) and (d).}}
 	\label{fig::Fig17_Ca_W_Di}
\end{figure}

In this section we  describe the effect of the capillary number on the main results. %
We consider a pair of cells with four different capillary numbers $C_a=\{5,10,25,100\}$ and flowing in channels of widths ranging from $2R_v$ to $20R_v$ ($\approx$ $6$ to $60$ $\mu m$ for RBCs). Figure \ref{fig::Fig17_Ca_W_Di} depicts the stationary interdistance as a function of the channel width for different capillary numbers (we do not show the full branch as before due to computational cost, and especially because we do not see any significant changes). %
Globally, the capillary number seems to not significantly affect the behavior of the pair.
\begin{figure}[h!]
 	\centering
 	\includegraphics[width = .65 \linewidth]{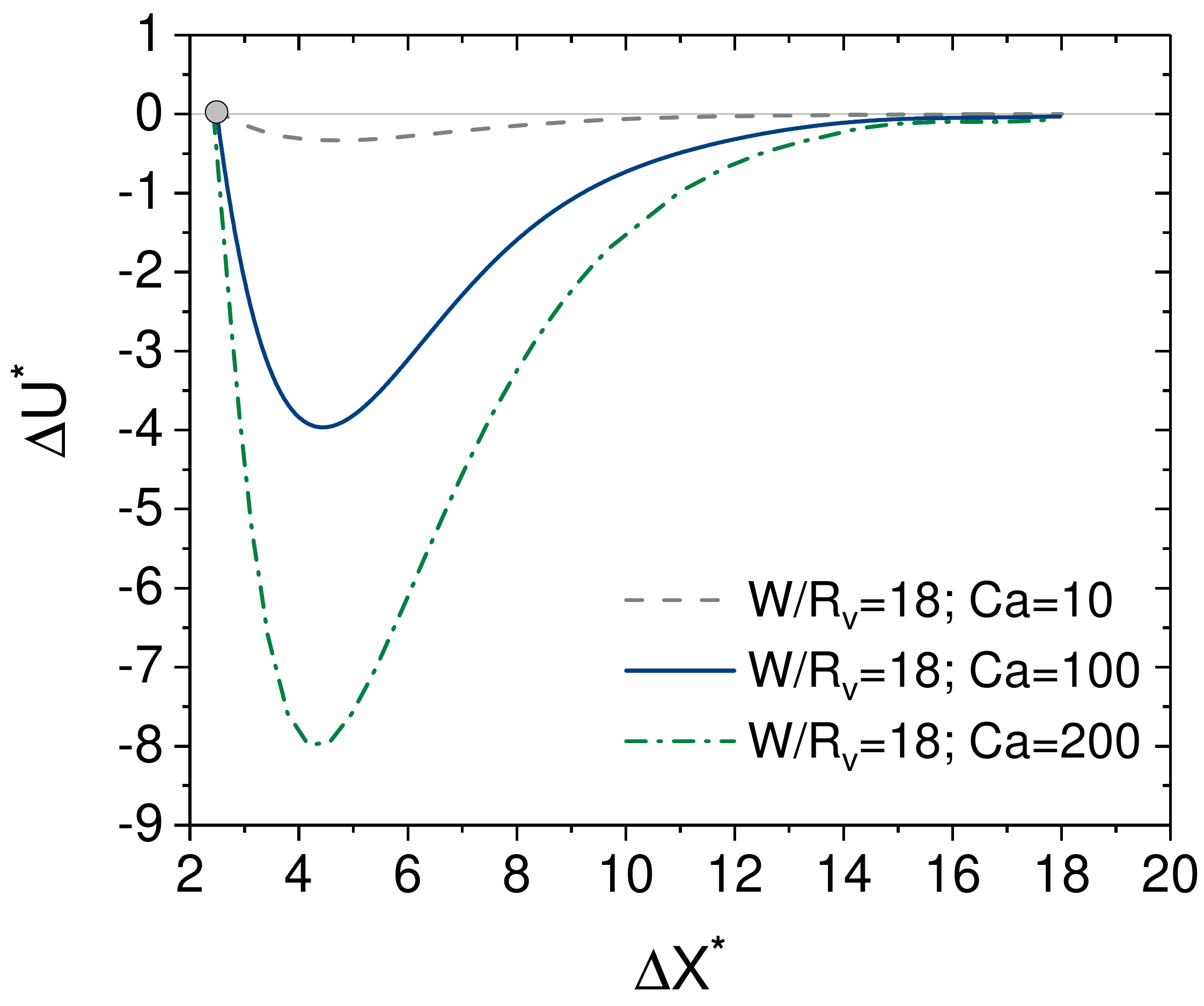}
 	\caption{ Relative velocity as a function of interdistance in weak confinement $W/R_v=18$ for different capillary numbers. The dashed (grey), solid (blue), and dash-dotted (green) lines are for $C_a=10$, $100$, and $200,$ respectively. %
 	}
 	\label{Vrelat_Ca100}
\end{figure}
For vesicles, higher values of $C_a$ can be reasonably reached, such as $2000$. For a vesicle of radius $20\; \mu m$, a channel radius of $200\;\mu m$, and a velocity
of $1\; cm/s$, one finds approximately $C_a\simeq 1600$. By assuming that the relative velocity scales with $C_a$ (as shown in  Figure \ref{Vrelat_Ca100}), one
finds that the relative velocity is of about $100 R_v/\tau\sim 1 R_v/s$, which is not devoid of experimental testability.
\section{Results for drops}
\label{sec:drops}
{Since most of the behaviors observed for vesicles are also observed for drops, we give only a brief discussion of the phase diagram for drop pairs. We have checked, as for vesicles, that initial shapes do not affect the final state. Drops can break up into several smaller drops when subject to a strong enough shear flow. The same effect occurs in Poiseuille flow. For this reason, it was important to choose the capillary number for which the drops remain stable. We chose $C_a=0.3.$ Figure \ref{drop_two} shows the steady state of a pair of drops and the corresponding flow lines. Unlike the vesicle case, drops exhibit recirculation zones inside in the form of two counter-rotating vortices. In addition, the velocity field at the interface is not constant along the contour as the drop interface is compressible.
\begin{figure}[h!]
 	\centering
 	\includegraphics[width = 1. \linewidth]{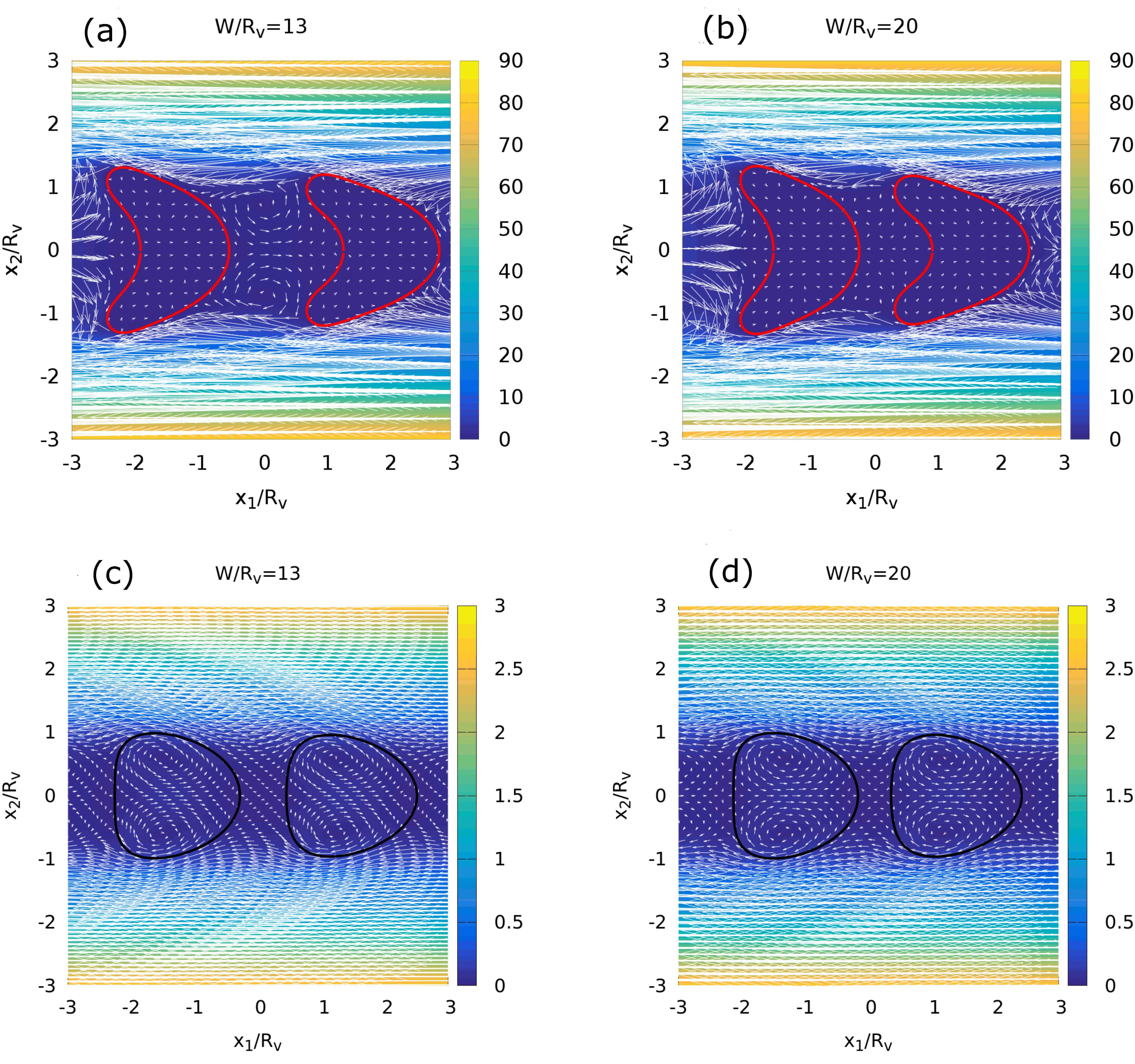}	
 	\caption{{The velocity field in the co-moving frame of a pair of drops (c and d) and vesicles (a and b) in a weakly confined flow ($W=20$ and $13$ $R_v$). $C_a=10$ for the vesicles and $C_a=0.3$ for the drops.}}
 	\label{drop_two}
\end{figure}

The question naturally arises of whether or not the apparent difference in the overall flow patterns as compared to vesicles results in different phase diagrams.  We have explored in a very systematic way the existence of branches of stationary solutions.
 The results are summarized in the phase diagram in Fig.\ref{diag_drop}. This phase diagram is strikingly similar to that obtained for vesicles (Fig. \ref{diag_full}). Not only is there a qualitative similarity but also the stationary interdistances obtained for drops in each confinement are close to those found for vesicles. The emergence of these strong similarities between the two systems points to the existence of a universal feature, where the details of the physical system do not matter too much. This lack of sensitivity of the hydrodynamic interaction to the mechanical properties of the interacting particles provides a prospect for further theoretical analysis.}
\begin{figure}[h!]
 	\centering
 	\includegraphics[width = .75 \linewidth]{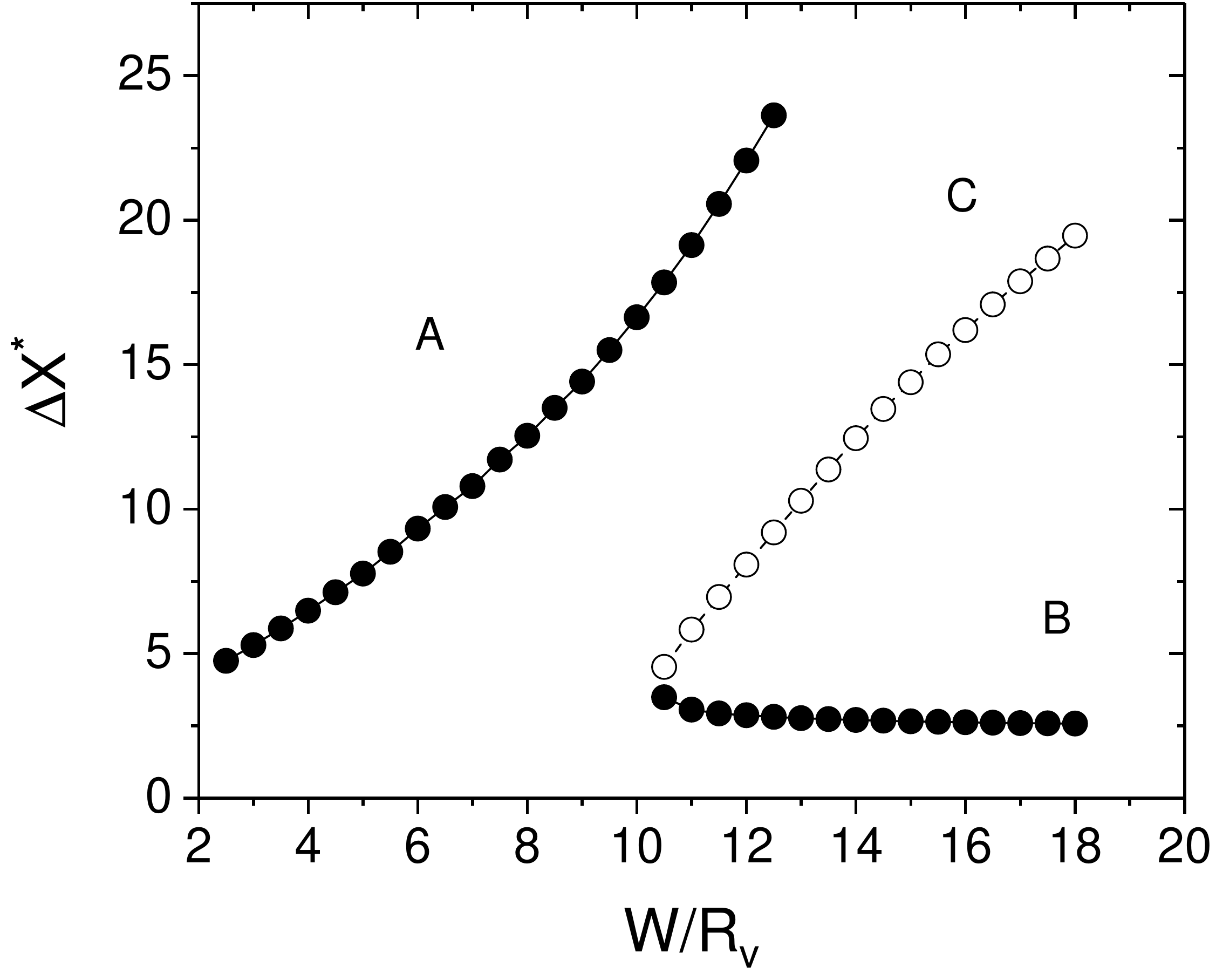}	
 	\caption{{Stationary interdistance for two drops as a function of the channel width. $C_a=0.3.$ }}
 	\label{diag_drop}
\end{figure}
\section{Conclusion}
\label{sec:conclusion}
{This study reports on a complex phase diagram regarding hydrodynamic interaction between two vesicles or two drops in a confined pressure-driven flow.
It is found that several branches of stationary solutions coexist. An interesting fact is that vesicles and drops behave in the same way, despite the different nature of the underlying physics between the two systems. This offers a useful basis for analytical modeling of the main features of the pairing process. As we have pointed out, for some branches of the phase diagram, the relative velocity of the pair of particles may become very small, in particular, due to the exponential decrease of the hydrodynamic interaction with interdistance. Thus other branches for longer interdistances may exist but be quite difficult to resolve numerically due to the smallness of interactions. While the existence of higher branches is an interesting fundamental question in itself, the very small amplitude of hydrodynamic interactions at long interdistances reduces the significance of this question for practical applications.

It would be interesting in the future to extend this study to the 3D case both for vesicles and RBCs. Another extension of this work is the analysis of many cells and more precisely how the stable size of a cluster evolves as a function of confinement. It has been reported in \cite{Giovanni2012} that in an unconfined Poiseuille flow, the cluster size depends on the flow strength: increasing the flow strength allows cluster of larger sizes to remain stable. It would be interesting to draw general conclusion about cluster stability in the presence of walls in order to complement the already existing literature on this topic \cite{freund2014numerical}.
A study of the stability of clusters would provide valuable information about the nature of modes that destabilize them, as studied recently in Ref. \cite{bryngelson2016capsule}. 
Another important question is to analyze the interplay between hydrodynamic interaction and that due to plasma proteins  (following our previous study  \cite{brust2014plasma}) and to study how the structure of the phase diagram reported here evolves in this case. }

\begin{acknowledgments}
We  acknowledge financial  support from the German Science Foundation research initiative SFB1027,
the German French University (DFH/UFA), CNES (Centre d'Etudes Spatiales) and ESA (European Space Agency).
\end{acknowledgments}

%\nocite{*}
%\bibliography{hydrobib}% Produces the bibliography via BibTeX.

\begin{thebibliography}{51}%
	\makeatletter
	\providecommand \@ifxundefined [1]{%
		\@ifx{#1\undefined}
	}%
	\providecommand \@ifnum [1]{%
		\ifnum #1\expandafter \@firstoftwo
		\else \expandafter \@secondoftwo
		\fi
	}%
	\providecommand \@ifx [1]{%
		\ifx #1\expandafter \@firstoftwo
		\else \expandafter \@secondoftwo
		\fi
	}%
	\providecommand \natexlab [1]{#1}%
	\providecommand \enquote  [1]{``#1''}%
	\providecommand \bibnamefont  [1]{#1}%
	\providecommand \bibfnamefont [1]{#1}%
	\providecommand \citenamefont [1]{#1}%
	\providecommand \href@noop [0]{\@secondoftwo}%
	\providecommand \href [0]{\begingroup \@sanitize@url \@href}%
	\providecommand \@href[1]{\@@startlink{#1}\@@href}%
	\providecommand \@@href[1]{\endgroup#1\@@endlink}%
	\providecommand \@sanitize@url [0]{\catcode `\\12\catcode `\$12\catcode
		`\&12\catcode `\#12\catcode `\^12\catcode `\_12\catcode `\%12\relax}%
	\providecommand \@@startlink[1]{}%
	\providecommand \@@endlink[0]{}%
	\providecommand \url  [0]{\begingroup\@sanitize@url \@url }%
	\providecommand \@url [1]{\endgroup\@href {#1}{\urlprefix }}%
	\providecommand \urlprefix  [0]{URL }%
	\providecommand \Eprint [0]{\href }%
	\providecommand \doibase [0]{http://dx.doi.org/}%
	\providecommand \selectlanguage [0]{\@gobble}%
	\providecommand \bibinfo  [0]{\@secondoftwo}%
	\providecommand \bibfield  [0]{\@secondoftwo}%
	\providecommand \translation [1]{[#1]}%
	\providecommand \BibitemOpen [0]{}%
	\providecommand \bibitemStop [0]{}%
	\providecommand \bibitemNoStop [0]{.\EOS\space}%
	\providecommand \EOS [0]{\spacefactor3000\relax}%
	\providecommand \BibitemShut  [1]{\csname bibitem#1\endcsname}%
	\let\auto@bib@innerbib\@empty
	%</preamble>
	\bibitem [{\citenamefont {Skalak}\ and\ \citenamefont
		{Branemark}(1969)}]{skalak69}%
	\BibitemOpen
	\bibfield  {author} {\bibinfo {author} {\bibfnamefont {R.}~\bibnamefont
			{Skalak}}\ and\ \bibinfo {author} {\bibfnamefont {P.~I.}\ \bibnamefont
			{Branemark}},\ }\href {\doibase 10.1126/science.164.3880.717} {\bibfield
		{journal} {\bibinfo  {journal} {Science}\ }\textbf {\bibinfo {volume}
			{164}},\ \bibinfo {pages} {717} (\bibinfo {year} {1969})}\BibitemShut
	{NoStop}%
	\bibitem [{\citenamefont {Gaehtgens}\ and\ \citenamefont
		{Schmid-Sch\"{o}nbein}(1982)}]{gaehtgens82}%
	\BibitemOpen
	\bibfield  {author} {\bibinfo {author} {\bibfnamefont {P.}~\bibnamefont
			{Gaehtgens}}\ and\ \bibinfo {author} {\bibfnamefont {H.}~\bibnamefont
			{Schmid-Sch\"{o}nbein}},\ }\href {\doibase 10.1007/BF00396444} {\bibfield
		{journal} {\bibinfo  {journal} {Naturwissenschaften}\ }\textbf {\bibinfo
			{volume} {69}},\ \bibinfo {pages} {294} (\bibinfo {year} {1982})}\BibitemShut
	{NoStop}%
	\bibitem [{\citenamefont {Tomaiuolo}\ \emph {et~al.}(2012)\citenamefont
		{Tomaiuolo}, \citenamefont {Lanotte}, \citenamefont {Ghigliotti},
		\citenamefont {Misbah},\ and\ \citenamefont {Guido}}]{tomaiuolo2012red}%
	\BibitemOpen
	\bibfield  {author} {\bibinfo {author} {\bibfnamefont {G.}~\bibnamefont
			{Tomaiuolo}}, \bibinfo {author} {\bibfnamefont {L.}~\bibnamefont {Lanotte}},
		\bibinfo {author} {\bibfnamefont {G.}~\bibnamefont {Ghigliotti}}, \bibinfo
		{author} {\bibfnamefont {C.}~\bibnamefont {Misbah}}, \ and\ \bibinfo {author}
		{\bibfnamefont {S.}~\bibnamefont {Guido}},\ }\href@noop {} {\bibfield
		{journal} {\bibinfo  {journal} {Physics of Fluids (1994-present)}\ }\textbf
		{\bibinfo {volume} {24}},\ \bibinfo {pages} {051903} (\bibinfo {year}
		{2012})}\BibitemShut {NoStop}%
	\bibitem [{\citenamefont {Brust}\ \emph {et~al.}(2014)\citenamefont {Brust},
		\citenamefont {Aouane}, \citenamefont {Thi{\'e}baud}, \citenamefont
		{Flormann}, \citenamefont {Verdier}, \citenamefont {Kaestner}, \citenamefont
		{Laschke}, \citenamefont {Selmi}, \citenamefont {Benyoussef}, \citenamefont
		{Podgorski} \emph {et~al.}}]{brust2014plasma}%
	\BibitemOpen
	\bibfield  {author} {\bibinfo {author} {\bibfnamefont {M.}~\bibnamefont
			{Brust}}, \bibinfo {author} {\bibfnamefont {O.}~\bibnamefont {Aouane}},
		\bibinfo {author} {\bibfnamefont {M.}~\bibnamefont {Thi{\'e}baud}}, \bibinfo
		{author} {\bibfnamefont {D.}~\bibnamefont {Flormann}}, \bibinfo {author}
		{\bibfnamefont {C.}~\bibnamefont {Verdier}}, \bibinfo {author} {\bibfnamefont
			{L.}~\bibnamefont {Kaestner}}, \bibinfo {author} {\bibfnamefont
			{M.}~\bibnamefont {Laschke}}, \bibinfo {author} {\bibfnamefont
			{H.}~\bibnamefont {Selmi}}, \bibinfo {author} {\bibfnamefont
			{A.}~\bibnamefont {Benyoussef}}, \bibinfo {author} {\bibfnamefont
			{T.}~\bibnamefont {Podgorski}},  \emph {et~al.},\ }\href@noop {} {\bibfield
		{journal} {\bibinfo  {journal} {Scientific reports}\ }\textbf {\bibinfo
			{volume} {4}} (\bibinfo {year} {2014})}\BibitemShut {NoStop}%
	\bibitem [{\citenamefont {Wang}\ and\ \citenamefont {Skalak}(1969)}]{wang1969}%
	\BibitemOpen
	\bibfield  {author} {\bibinfo {author} {\bibfnamefont {H.}~\bibnamefont
			{Wang}}\ and\ \bibinfo {author} {\bibfnamefont {R.}~\bibnamefont {Skalak}},\
	}\href {\doibase 10.1017/S002211206900005X} {\bibfield  {journal} {\bibinfo
			{journal} {Journal of Fluid Mechanics}\ }\textbf {\bibinfo {volume} {38}},\
		\bibinfo {pages} {75} (\bibinfo {year} {1969})}\BibitemShut {NoStop}%
	\bibitem [{\citenamefont {Leichtberg}\ \emph {et~al.}(1976)\citenamefont
		{Leichtberg}, \citenamefont {Pfeffer},\ and\ \citenamefont
		{Weinbaum}}]{leichtberg1976stokes}%
	\BibitemOpen
	\bibfield  {author} {\bibinfo {author} {\bibfnamefont {S.}~\bibnamefont
			{Leichtberg}}, \bibinfo {author} {\bibfnamefont {R.}~\bibnamefont {Pfeffer}},
		\ and\ \bibinfo {author} {\bibfnamefont {S.}~\bibnamefont {Weinbaum}},\
	}\href@noop {} {\bibfield  {journal} {\bibinfo  {journal} {International
				Journal of Multiphase Flow}\ }\textbf {\bibinfo {volume} {3}},\ \bibinfo
		{pages} {147} (\bibinfo {year} {1976})}\BibitemShut {NoStop}%
	\bibitem [{\citenamefont {Cui}\ \emph {et~al.}(2002)\citenamefont {Cui},
		\citenamefont {Diamant},\ and\ \citenamefont {Lin}}]{Cui2002}%
	\BibitemOpen
	\bibfield  {author} {\bibinfo {author} {\bibfnamefont {B.}~\bibnamefont
			{Cui}}, \bibinfo {author} {\bibfnamefont {H.}~\bibnamefont {Diamant}}, \ and\
		\bibinfo {author} {\bibfnamefont {B.}~\bibnamefont {Lin}},\ }\href {\doibase
		10.1103/PhysRevLett.89.188302} {\bibfield  {journal} {\bibinfo  {journal}
			{Phys. Rev. Lett.}\ }\textbf {\bibinfo {volume} {89}},\ \bibinfo {pages}
		{188302} (\bibinfo {year} {2002})}\BibitemShut {NoStop}%
	\bibitem [{\citenamefont {Cui}\ \emph {et~al.}(2004)\citenamefont {Cui},
		\citenamefont {Diamant}, \citenamefont {Lin},\ and\ \citenamefont
		{Rice}}]{Cui2004}%
	\BibitemOpen
	\bibfield  {author} {\bibinfo {author} {\bibfnamefont {B.}~\bibnamefont
			{Cui}}, \bibinfo {author} {\bibfnamefont {H.}~\bibnamefont {Diamant}},
		\bibinfo {author} {\bibfnamefont {B.}~\bibnamefont {Lin}}, \ and\ \bibinfo
		{author} {\bibfnamefont {S.~A.}~\bibnamefont {Rice}},\ }\href {\doibase
		10.1103/PhysRevLett.92.258301} {\bibfield  {journal} {\bibinfo  {journal}
			{Phys. Rev. Lett.}\ }\textbf {\bibinfo {volume} {92}},\ \bibinfo {pages}
		{258301} (\bibinfo {year} {2004})}\BibitemShut {NoStop}%
	\bibitem [{\citenamefont {Diamant}\ \emph {et~al.}(2005)\citenamefont
		{Diamant}, \citenamefont {Cui}, \citenamefont {Lin},\ and\ \citenamefont
		{Rice}}]{Diamant2005}%
	\BibitemOpen
	\bibfield  {author} {\bibinfo {author} {\bibfnamefont {H.}~\bibnamefont
			{Diamant}}, \bibinfo {author} {\bibfnamefont {B.}~\bibnamefont {Cui}},
		\bibinfo {author} {\bibfnamefont {B.}~\bibnamefont {Lin}}, \ and\ \bibinfo
		{author} {\bibfnamefont {S.~A.}~\bibnamefont {Rice}},\ }\href
	{http://stacks.iop.org/0953-8984/17/i=31/a=003} {\bibfield  {journal}
		{\bibinfo  {journal} {Journal of Physics: Condensed Matter}\ }\textbf
		{\bibinfo {volume} {17}},\ \bibinfo {pages} {S2787} (\bibinfo {year}
		{2005})}\BibitemShut {NoStop}%
	\bibitem [{\citenamefont {Beatus}\ \emph {et~al.}(2006)\citenamefont {Beatus},
		\citenamefont {Tlusty},\ and\ \citenamefont {Bar-Ziv}}]{beatus2006phonons}%
	\BibitemOpen
	\bibfield  {author} {\bibinfo {author} {\bibfnamefont {T.}~\bibnamefont
			{Beatus}}, \bibinfo {author} {\bibfnamefont {T.}~\bibnamefont {Tlusty}}, \
		and\ \bibinfo {author} {\bibfnamefont {R.}~\bibnamefont {Bar-Ziv}},\
	}\href@noop {} {\bibfield  {journal} {\bibinfo  {journal} {Nature Physics}\
		}\textbf {\bibinfo {volume} {2}},\ \bibinfo {pages} {743} (\bibinfo {year}
		{2006})}\BibitemShut {NoStop}%
	\bibitem [{\citenamefont {Beatus}\ \emph {et~al.}(2007)\citenamefont {Beatus},
		\citenamefont {Bar-Ziv},\ and\ \citenamefont {Tlusty}}]{beatus2007anomalous}%
	\BibitemOpen
	\bibfield  {author} {\bibinfo {author} {\bibfnamefont {T.}~\bibnamefont
			{Beatus}}, \bibinfo {author} {\bibfnamefont {R.}~\bibnamefont {Bar-Ziv}}, \
		and\ \bibinfo {author} {\bibfnamefont {T.}~\bibnamefont {Tlusty}},\
	}\href@noop {} {\bibfield  {journal} {\bibinfo  {journal} {Physical review
				letters}\ }\textbf {\bibinfo {volume} {99}},\ \bibinfo {pages} {124502}
		(\bibinfo {year} {2007})}\BibitemShut {NoStop}%
	\bibitem [{\citenamefont {Shani}\ \emph {et~al.}(2014)\citenamefont {Shani},
		\citenamefont {Beatus}, \citenamefont {Bar-Ziv},\ and\ \citenamefont
		{Tlusty}}]{shani2014long}%
	\BibitemOpen
	\bibfield  {author} {\bibinfo {author} {\bibfnamefont {I.}~\bibnamefont
			{Shani}}, \bibinfo {author} {\bibfnamefont {T.}~\bibnamefont {Beatus}},
		\bibinfo {author} {\bibfnamefont {R.~H.}\ \bibnamefont {Bar-Ziv}}, \ and\
		\bibinfo {author} {\bibfnamefont {T.}~\bibnamefont {Tlusty}},\ }\href@noop {}
	{\bibfield  {journal} {\bibinfo  {journal} {Nature Physics}\ }\textbf
		{\bibinfo {volume} {10}},\ \bibinfo {pages} {140} (\bibinfo {year}
		{2014})}\BibitemShut {NoStop}%
	\bibitem [{\citenamefont {Janssen}\ \emph {et~al.}(2012)\citenamefont
		{Janssen}, \citenamefont {Baron}, \citenamefont {Anderson}, \citenamefont
		{Blawzdziewicz}, \citenamefont {Loewenberg},\ and\ \citenamefont
		{Wajnryb}}]{janssen2012collective}%
	\BibitemOpen
	\bibfield  {author} {\bibinfo {author} {\bibfnamefont {P.}~\bibnamefont
			{Janssen}}, \bibinfo {author} {\bibfnamefont {M.}~\bibnamefont {Baron}},
		\bibinfo {author} {\bibfnamefont {P.}~\bibnamefont {Anderson}}, \bibinfo
		{author} {\bibfnamefont {J.}~\bibnamefont {Blawzdziewicz}}, \bibinfo {author}
		{\bibfnamefont {M.}~\bibnamefont {Loewenberg}}, \ and\ \bibinfo {author}
		{\bibfnamefont {E.}~\bibnamefont {Wajnryb}},\ }\href@noop {} {\bibfield
		{journal} {\bibinfo  {journal} {Soft Matter}\ }\textbf {\bibinfo {volume}
			{8}},\ \bibinfo {pages} {7495} (\bibinfo {year} {2012})}\BibitemShut
	{NoStop}%
	\bibitem [{\citenamefont {Breyiannis}\ and\ \citenamefont
		{Pozrikidis}(2000)}]{breyiannis2000simple}%
	\BibitemOpen
	\bibfield  {author} {\bibinfo {author} {\bibfnamefont {G.}~\bibnamefont
			{Breyiannis}}\ and\ \bibinfo {author} {\bibfnamefont {C.}~\bibnamefont
			{Pozrikidis}},\ }\href@noop {} {\bibfield  {journal} {\bibinfo  {journal}
			{Theoretical and Computational Fluid Dynamics}\ }\textbf {\bibinfo {volume}
			{13}},\ \bibinfo {pages} {327} (\bibinfo {year} {2000})}\BibitemShut
	{NoStop}%
	\bibitem [{\citenamefont {Liu}\ and\ \citenamefont
		{Liu}(2006)}]{liu2006rheology}%
	\BibitemOpen
	\bibfield  {author} {\bibinfo {author} {\bibfnamefont {Y.}~\bibnamefont
			{Liu}}\ and\ \bibinfo {author} {\bibfnamefont {W.~K.}\ \bibnamefont {Liu}},\
	}\href@noop {} {\bibfield  {journal} {\bibinfo  {journal} {Journal of
				Computational Physics}\ }\textbf {\bibinfo {volume} {220}},\ \bibinfo {pages}
		{139} (\bibinfo {year} {2006})}\BibitemShut {NoStop}%
	\bibitem [{\citenamefont {Secomb}\ \emph {et~al.}(2007)\citenamefont {Secomb},
		\citenamefont {Styp-Rekowska},\ and\ \citenamefont {Pries}}]{secomb2007two}%
	\BibitemOpen
	\bibfield  {author} {\bibinfo {author} {\bibfnamefont {T.~W.}\ \bibnamefont
			{Secomb}}, \bibinfo {author} {\bibfnamefont {B.}~\bibnamefont
			{Styp-Rekowska}}, \ and\ \bibinfo {author} {\bibfnamefont {A.~R.}\
			\bibnamefont {Pries}},\ }\href@noop {} {\bibfield  {journal} {\bibinfo
			{journal} {Annals of biomedical engineering}\ }\textbf {\bibinfo {volume}
			{35}},\ \bibinfo {pages} {755} (\bibinfo {year} {2007})}\BibitemShut
	{NoStop}%
	\bibitem [{\citenamefont {Dupin}\ \emph {et~al.}(2007)\citenamefont {Dupin},
		\citenamefont {Halliday}, \citenamefont {Care}, \citenamefont {Alboul},\ and\
		\citenamefont {Munn}}]{dupin2007modeling}%
	\BibitemOpen
	\bibfield  {author} {\bibinfo {author} {\bibfnamefont {M.~M.}\ \bibnamefont
			{Dupin}}, \bibinfo {author} {\bibfnamefont {I.}~\bibnamefont {Halliday}},
		\bibinfo {author} {\bibfnamefont {C.~M.}\ \bibnamefont {Care}}, \bibinfo
		{author} {\bibfnamefont {L.}~\bibnamefont {Alboul}}, \ and\ \bibinfo {author}
		{\bibfnamefont {L.~L.}\ \bibnamefont {Munn}},\ }\href@noop {} {\bibfield
		{journal} {\bibinfo  {journal} {Physical Review E}\ }\textbf {\bibinfo
			{volume} {75}},\ \bibinfo {pages} {066707} (\bibinfo {year}
		{2007})}\BibitemShut {NoStop}%
	\bibitem [{\citenamefont {Vlahovska}\ \emph {et~al.}(2009)\citenamefont
		{Vlahovska}, \citenamefont {Podgorski},\ and\ \citenamefont
		{Misbah}}]{vlahovska2009vesicles}%
	\BibitemOpen
	\bibfield  {author} {\bibinfo {author} {\bibfnamefont {P.~M.}\ \bibnamefont
			{Vlahovska}}, \bibinfo {author} {\bibfnamefont {T.}~\bibnamefont
			{Podgorski}}, \ and\ \bibinfo {author} {\bibfnamefont {C.}~\bibnamefont
			{Misbah}},\ }\href@noop {} {\bibfield  {journal} {\bibinfo  {journal}
			{Comptes Rendus Physique}\ }\textbf {\bibinfo {volume} {10}},\ \bibinfo
		{pages} {775} (\bibinfo {year} {2009})}\BibitemShut {NoStop}%
	\bibitem [{\citenamefont {Doddi}\ and\ \citenamefont
		{Bagchi}(2009)}]{doddi2009three}%
	\BibitemOpen
	\bibfield  {author} {\bibinfo {author} {\bibfnamefont {S.~K.}\ \bibnamefont
			{Doddi}}\ and\ \bibinfo {author} {\bibfnamefont {P.}~\bibnamefont {Bagchi}},\
	}\href@noop {} {\bibfield  {journal} {\bibinfo  {journal} {Physical Review
				E}\ }\textbf {\bibinfo {volume} {79}},\ \bibinfo {pages} {046318} (\bibinfo
		{year} {2009})}\BibitemShut {NoStop}%
	\bibitem [{\citenamefont {McWhirter}\ \emph {et~al.}(2009)\citenamefont
		{McWhirter}, \citenamefont {Noguchi},\ and\ \citenamefont
		{Gompper}}]{mcwhirter2009flow}%
	\BibitemOpen
	\bibfield  {author} {\bibinfo {author} {\bibfnamefont {J.~L.}\ \bibnamefont
			{McWhirter}}, \bibinfo {author} {\bibfnamefont {H.}~\bibnamefont {Noguchi}},
		\ and\ \bibinfo {author} {\bibfnamefont {G.}~\bibnamefont {Gompper}},\
	}\href@noop {} {\bibfield  {journal} {\bibinfo  {journal} {Proceedings of the
				National Academy of Sciences}\ }\textbf {\bibinfo {volume} {106}},\ \bibinfo
		{pages} {6039} (\bibinfo {year} {2009})}\BibitemShut {NoStop}%
	\bibitem [{\citenamefont {Veerapaneni}\ \emph {et~al.}(2009)\citenamefont
		{Veerapaneni}, \citenamefont {Gueyffier}, \citenamefont {Zorin},\ and\
		\citenamefont {Biros}}]{veerapaneni2009boundary}%
	\BibitemOpen
	\bibfield  {author} {\bibinfo {author} {\bibfnamefont {S.~K.}\ \bibnamefont
			{Veerapaneni}}, \bibinfo {author} {\bibfnamefont {D.}~\bibnamefont
			{Gueyffier}}, \bibinfo {author} {\bibfnamefont {D.}~\bibnamefont {Zorin}}, \
		and\ \bibinfo {author} {\bibfnamefont {G.}~\bibnamefont {Biros}},\
	}\href@noop {} {\bibfield  {journal} {\bibinfo  {journal} {Journal of
				Computational Physics}\ }\textbf {\bibinfo {volume} {228}},\ \bibinfo {pages}
		{2334} (\bibinfo {year} {2009})}\BibitemShut {NoStop}%
	\bibitem [{\citenamefont {Zhao}\ \emph {et~al.}(2010)\citenamefont {Zhao},
		\citenamefont {Isfahani}, \citenamefont {Olson},\ and\ \citenamefont
		{Freund}}]{zhao2010spectral}%
	\BibitemOpen
	\bibfield  {author} {\bibinfo {author} {\bibfnamefont {H.}~\bibnamefont
			{Zhao}}, \bibinfo {author} {\bibfnamefont {A.~H.}\ \bibnamefont {Isfahani}},
		\bibinfo {author} {\bibfnamefont {L.~N.}\ \bibnamefont {Olson}}, \ and\
		\bibinfo {author} {\bibfnamefont {J.~B.}\ \bibnamefont {Freund}},\
	}\href@noop {} {\bibfield  {journal} {\bibinfo  {journal} {Journal of
				Computational Physics}\ }\textbf {\bibinfo {volume} {229}},\ \bibinfo {pages}
		{3726} (\bibinfo {year} {2010})}\BibitemShut {NoStop}%
	\bibitem [{\citenamefont {Zhao}\ and\ \citenamefont
		{Shaqfeh}(2011)}]{zhao2011shear}%
	\BibitemOpen
	\bibfield  {author} {\bibinfo {author} {\bibfnamefont {H.}~\bibnamefont
			{Zhao}}\ and\ \bibinfo {author} {\bibfnamefont {E.~S.~G.}\ \bibnamefont
			{Shaqfeh}},\ }\href@noop {} {\bibfield  {journal} {\bibinfo  {journal}
			{Physical Review E}\ }\textbf {\bibinfo {volume} {83}},\ \bibinfo {pages}
		{061924} (\bibinfo {year} {2011})}\BibitemShut {NoStop}%
	\bibitem [{\citenamefont {McWhirter}\ \emph {et~al.}(2011)\citenamefont
		{McWhirter}, \citenamefont {Noguchi},\ and\ \citenamefont
		{Gompper}}]{mcwhirter2011deformation}%
	\BibitemOpen
	\bibfield  {author} {\bibinfo {author} {\bibfnamefont {J.}~\bibnamefont
			{McWhirter}}, \bibinfo {author} {\bibfnamefont {H.}~\bibnamefont {Noguchi}},
		\ and\ \bibinfo {author} {\bibfnamefont {G.}~\bibnamefont {Gompper}},\
	}\href@noop {} {\bibfield  {journal} {\bibinfo  {journal} {Soft Matter}\
		}\textbf {\bibinfo {volume} {7}},\ \bibinfo {pages} {10967} (\bibinfo {year}
		{2011})}\BibitemShut {NoStop}%
	\bibitem [{\citenamefont {Freund}\ and\ \citenamefont
		{Orescanin}(2011)}]{freund2011cellular}%
	\BibitemOpen
	\bibfield  {author} {\bibinfo {author} {\bibfnamefont {J.~B.}\ \bibnamefont
			{Freund}}\ and\ \bibinfo {author} {\bibfnamefont {M.}~\bibnamefont
			{Orescanin}},\ }\href@noop {} {\bibfield  {journal} {\bibinfo  {journal}
			{Journal of Fluid Mechanics}\ }\textbf {\bibinfo {volume} {671}},\ \bibinfo
		{pages} {466} (\bibinfo {year} {2011})}\BibitemShut {NoStop}%
	\bibitem [{\citenamefont {Kr{\"u}ger}\ \emph {et~al.}(2011)\citenamefont
		{Kr{\"u}ger}, \citenamefont {Varnik},\ and\ \citenamefont
		{Raabe}}]{kruger2011particle}%
	\BibitemOpen
	\bibfield  {author} {\bibinfo {author} {\bibfnamefont {T.}~\bibnamefont
			{Kr{\"u}ger}}, \bibinfo {author} {\bibfnamefont {F.}~\bibnamefont {Varnik}},
		\ and\ \bibinfo {author} {\bibfnamefont {D.}~\bibnamefont {Raabe}},\
	}\href@noop {} {\bibfield  {journal} {\bibinfo  {journal} {Philosophical
				Transactions of the Royal Society of London A: Mathematical, Physical and
				Engineering Sciences}\ }\textbf {\bibinfo {volume} {369}},\ \bibinfo {pages}
		{2414} (\bibinfo {year} {2011})}\BibitemShut {NoStop}%
	\bibitem [{\citenamefont {Fedosov}\ \emph {et~al.}(2011)\citenamefont
		{Fedosov}, \citenamefont {Pan}, \citenamefont {Caswell}, \citenamefont
		{Gompper},\ and\ \citenamefont {Karniadakis}}]{fedosov2011predicting}%
	\BibitemOpen
	\bibfield  {author} {\bibinfo {author} {\bibfnamefont {D.~A.}\ \bibnamefont
			{Fedosov}}, \bibinfo {author} {\bibfnamefont {W.}~\bibnamefont {Pan}},
		\bibinfo {author} {\bibfnamefont {B.}~\bibnamefont {Caswell}}, \bibinfo
		{author} {\bibfnamefont {G.}~\bibnamefont {Gompper}}, \ and\ \bibinfo
		{author} {\bibfnamefont {G.~E.}\ \bibnamefont {Karniadakis}},\ }\href@noop {}
	{\bibfield  {journal} {\bibinfo  {journal} {Proceedings of the National
				Academy of Sciences}\ }\textbf {\bibinfo {volume} {108}},\ \bibinfo {pages}
		{11772} (\bibinfo {year} {2011})}\BibitemShut {NoStop}%
	\bibitem [{\citenamefont {Alizadehrad}\ \emph {et~al.}(2012)\citenamefont
		{Alizadehrad}, \citenamefont {Imai}, \citenamefont {Nakaaki}, \citenamefont
		{Ishikawa},\ and\ \citenamefont {Yamaguchi}}]{alizadehrad2012quantification}%
	\BibitemOpen
	\bibfield  {author} {\bibinfo {author} {\bibfnamefont {D.}~\bibnamefont
			{Alizadehrad}}, \bibinfo {author} {\bibfnamefont {Y.}~\bibnamefont {Imai}},
		\bibinfo {author} {\bibfnamefont {K.}~\bibnamefont {Nakaaki}}, \bibinfo
		{author} {\bibfnamefont {T.}~\bibnamefont {Ishikawa}}, \ and\ \bibinfo
		{author} {\bibfnamefont {T.}~\bibnamefont {Yamaguchi}},\ }\href@noop {}
	{\bibfield  {journal} {\bibinfo  {journal} {Journal of biomechanics}\
		}\textbf {\bibinfo {volume} {45}},\ \bibinfo {pages} {2684} (\bibinfo {year}
		{2012})}\BibitemShut {NoStop}%
	\bibitem [{\citenamefont {Reasor}\ \emph {et~al.}(2012)\citenamefont {Reasor},
		\citenamefont {Clausen},\ and\ \citenamefont {Aidun}}]{reasor2012coupling}%
	\BibitemOpen
	\bibfield  {author} {\bibinfo {author} {\bibfnamefont {D.~A.}\ \bibnamefont
			{Reasor}}, \bibinfo {author} {\bibfnamefont {J.~R.}\ \bibnamefont {Clausen}},
		\ and\ \bibinfo {author} {\bibfnamefont {C.~K.}\ \bibnamefont {Aidun}},\
	}\href@noop {} {\bibfield  {journal} {\bibinfo  {journal} {International
				Journal for Numerical Methods in Fluids}\ }\textbf {\bibinfo {volume} {68}},\
		\bibinfo {pages} {767} (\bibinfo {year} {2012})}\BibitemShut {NoStop}%
	\bibitem [{\citenamefont {Zhao}\ and\ \citenamefont
		{Shaqfeh}(2013)}]{zhao2013dynamics}%
	\BibitemOpen
	\bibfield  {author} {\bibinfo {author} {\bibfnamefont {H.}~\bibnamefont
			{Zhao}}\ and\ \bibinfo {author} {\bibfnamefont {E.~S.}\ \bibnamefont
			{Shaqfeh}},\ }\href@noop {} {\bibfield  {journal} {\bibinfo  {journal}
			{Journal of Fluid Mechanics}\ }\textbf {\bibinfo {volume} {725}},\ \bibinfo
		{pages} {709} (\bibinfo {year} {2013})}\BibitemShut {NoStop}%
	\bibitem [{\citenamefont {Thi\'ebaud}\ and\ \citenamefont
		{Misbah}(2013)}]{marine2013wallsgreenfunction}%
	\BibitemOpen
	\bibfield  {author} {\bibinfo {author} {\bibfnamefont {M.}~\bibnamefont
			{Thi\'ebaud}}\ and\ \bibinfo {author} {\bibfnamefont {C.}~\bibnamefont
			{Misbah}},\ }\href {\doibase 10.1103/PhysRevE.88.062707} {\bibfield
		{journal} {\bibinfo  {journal} {Phys. Rev. E}\ }\textbf {\bibinfo {volume}
			{88}},\ \bibinfo {pages} {062707} (\bibinfo {year} {2013})}\BibitemShut
	{NoStop}%
	\bibitem [{\citenamefont {Freund}(2014)}]{freund2014numerical}%
	\BibitemOpen
	\bibfield  {author} {\bibinfo {author} {\bibfnamefont {J.~B.}\ \bibnamefont
			{Freund}},\ }\href@noop {} {\bibfield  {journal} {\bibinfo  {journal} {Annual
				review of fluid mechanics}\ }\textbf {\bibinfo {volume} {46}},\ \bibinfo
		{pages} {67} (\bibinfo {year} {2014})}\BibitemShut {NoStop}%
	\bibitem [{\citenamefont {Matsunaga}\ \emph {et~al.}(2015)\citenamefont
		{Matsunaga}, \citenamefont {Imai}, \citenamefont {Yamaguchi},\ and\
		\citenamefont {Ishikawa}}]{matsunaga2015rheology}%
	\BibitemOpen
	\bibfield  {author} {\bibinfo {author} {\bibfnamefont {D.}~\bibnamefont
			{Matsunaga}}, \bibinfo {author} {\bibfnamefont {Y.}~\bibnamefont {Imai}},
		\bibinfo {author} {\bibfnamefont {T.}~\bibnamefont {Yamaguchi}}, \ and\
		\bibinfo {author} {\bibfnamefont {T.}~\bibnamefont {Ishikawa}},\ }\href@noop
	{} {\bibfield  {journal} {\bibinfo  {journal} {Journal of Fluid Mechanics}\
		}\textbf {\bibinfo {volume} {786}},\ \bibinfo {pages} {110} (\bibinfo {year}
		{2015})}\BibitemShut {NoStop}%
	\bibitem [{\citenamefont {Bryngelson}\ and\ \citenamefont
		{Freund}(2016)}]{bryngelson2016capsule}%
	\BibitemOpen
	\bibfield  {author} {\bibinfo {author} {\bibfnamefont {S.~H.}\ \bibnamefont
			{Bryngelson}}\ and\ \bibinfo {author} {\bibfnamefont {J.~B.}\ \bibnamefont
			{Freund}},\ }\href@noop {} {\bibfield  {journal} {\bibinfo  {journal}
			{Physical Review Fluids}\ }\textbf {\bibinfo {volume} {1}},\ \bibinfo {pages}
		{033201} (\bibinfo {year} {2016})}\BibitemShut {NoStop}%
	\bibitem [{\citenamefont {Ghigliotti}\ \emph {et~al.}(2012)\citenamefont
		{Ghigliotti}, \citenamefont {Selmi}, \citenamefont {Asmi},\ and\
		\citenamefont {Misbah}}]{Giovanni2012}%
	\BibitemOpen
	\bibfield  {author} {\bibinfo {author} {\bibfnamefont {G.}~\bibnamefont
			{Ghigliotti}}, \bibinfo {author} {\bibfnamefont {H.}~\bibnamefont {Selmi}},
		\bibinfo {author} {\bibfnamefont {L.~E.}\ \bibnamefont {Asmi}}, \ and\
		\bibinfo {author} {\bibfnamefont {C.}~\bibnamefont {Misbah}},\ }\href
	{\doibase http://dx.doi.org/10.1063/1.4757394} {\bibfield  {journal}
		{\bibinfo  {journal} {Physics of Fluids (1994-present)}\ }\textbf {\bibinfo
			{volume} {24}},\ \bibinfo {eid} {101901} (\bibinfo {year}
		{2012})}\BibitemShut {NoStop}%
	\bibitem [{\citenamefont {Misbah}(2006)}]{misbah2006}%
	\BibitemOpen
	\bibfield  {author} {\bibinfo {author} {\bibfnamefont {C.}~\bibnamefont
			{Misbah}},\ }\href {\doibase 10.1103/PhysRevLett.96.028104} {\bibfield
		{journal} {\bibinfo  {journal} {Phys. Rev. Lett.}\ }\textbf {\bibinfo
			{volume} {96}},\ \bibinfo {pages} {028104} (\bibinfo {year}
		{2006})}\BibitemShut {NoStop}%
	\bibitem [{\citenamefont {Noguchi}\ and\ \citenamefont
		{Gompper}(2007)}]{noguchi2007swinging}%
	\BibitemOpen
	\bibfield  {author} {\bibinfo {author} {\bibfnamefont {H.}~\bibnamefont
			{Noguchi}}\ and\ \bibinfo {author} {\bibfnamefont {G.}~\bibnamefont
			{Gompper}},\ }\href@noop {} {\bibfield  {journal} {\bibinfo  {journal}
			{Physical review letters}\ }\textbf {\bibinfo {volume} {98}},\ \bibinfo
		{pages} {128103} (\bibinfo {year} {2007})}\BibitemShut {NoStop}%
	\bibitem [{\citenamefont {Farutin}\ \emph {et~al.}(2012)\citenamefont
		{Farutin}, \citenamefont {Aouane},\ and\ \citenamefont
		{Misbah}}]{farutin2012vesicle}%
	\BibitemOpen
	\bibfield  {author} {\bibinfo {author} {\bibfnamefont {A.}~\bibnamefont
			{Farutin}}, \bibinfo {author} {\bibfnamefont {O.}~\bibnamefont {Aouane}}, \
		and\ \bibinfo {author} {\bibfnamefont {C.}~\bibnamefont {Misbah}},\
	}\href@noop {} {\bibfield  {journal} {\bibinfo  {journal} {Physical Review
				E}\ }\textbf {\bibinfo {volume} {85}},\ \bibinfo {pages} {061922} (\bibinfo
		{year} {2012})}\BibitemShut {NoStop}%
	\bibitem [{\citenamefont {Aouane}\ \emph {et~al.}(2014)\citenamefont {Aouane},
		\citenamefont {Thi\'ebaud}, \citenamefont {Benyoussef}, \citenamefont
		{Wagner},\ and\ \citenamefont {Misbah}}]{Othmane2014}%
	\BibitemOpen
	\bibfield  {author} {\bibinfo {author} {\bibfnamefont {O.}~\bibnamefont
			{Aouane}}, \bibinfo {author} {\bibfnamefont {M.}~\bibnamefont {Thi\'ebaud}},
		\bibinfo {author} {\bibfnamefont {A.}~\bibnamefont {Benyoussef}}, \bibinfo
		{author} {\bibfnamefont {C.}~\bibnamefont {Wagner}}, \ and\ \bibinfo {author}
		{\bibfnamefont {C.}~\bibnamefont {Misbah}},\ }\href {\doibase
		10.1103/PhysRevE.90.033011} {\bibfield  {journal} {\bibinfo  {journal} {Phys.
				Rev. E}\ }\textbf {\bibinfo {volume} {90}},\ \bibinfo {pages} {033011}
		(\bibinfo {year} {2014})}\BibitemShut {NoStop}%
	\bibitem [{\citenamefont {Ramanujan}\ and\ \citenamefont
		{Pozrikidis}(1998)}]{ramanujan1998deformation}%
	\BibitemOpen
	\bibfield  {author} {\bibinfo {author} {\bibfnamefont {S.}~\bibnamefont
			{Ramanujan}}\ and\ \bibinfo {author} {\bibfnamefont {C.}~\bibnamefont
			{Pozrikidis}},\ }\href@noop {} {\bibfield  {journal} {\bibinfo  {journal}
			{Journal of Fluid Mechanics}\ }\textbf {\bibinfo {volume} {361}},\ \bibinfo
		{pages} {117} (\bibinfo {year} {1998})}\BibitemShut {NoStop}%
	\bibitem [{\citenamefont {Lac}\ and\ \citenamefont
		{Barth{\`e}s-Biesel}(2005)}]{lac2005deformation}%
	\BibitemOpen
	\bibfield  {author} {\bibinfo {author} {\bibfnamefont {E.}~\bibnamefont
			{Lac}}\ and\ \bibinfo {author} {\bibfnamefont {D.}~\bibnamefont
			{Barth{\`e}s-Biesel}},\ }\href@noop {} {\bibfield  {journal} {\bibinfo
			{journal} {Physics of Fluids (1994-present)}\ }\textbf {\bibinfo {volume}
			{17}},\ \bibinfo {pages} {072105} (\bibinfo {year} {2005})}\BibitemShut
	{NoStop}%
	\bibitem [{\citenamefont {Kessler}\ \emph {et~al.}(2008)\citenamefont
		{Kessler}, \citenamefont {Finken},\ and\ \citenamefont
		{Seifert}}]{kessler2008swinging}%
	\BibitemOpen
	\bibfield  {author} {\bibinfo {author} {\bibfnamefont {S.}~\bibnamefont
			{Kessler}}, \bibinfo {author} {\bibfnamefont {R.}~\bibnamefont {Finken}}, \
		and\ \bibinfo {author} {\bibfnamefont {U.}~\bibnamefont {Seifert}},\
	}\href@noop {} {\bibfield  {journal} {\bibinfo  {journal} {Journal of Fluid
				Mechanics}\ }\textbf {\bibinfo {volume} {605}},\ \bibinfo {pages} {207}
		(\bibinfo {year} {2008})}\BibitemShut {NoStop}%
	\bibitem [{\citenamefont {Helfrich}(1973)}]{helfrich1973}%
	\BibitemOpen
	\bibfield  {author} {\bibinfo {author} {\bibfnamefont {W.}~\bibnamefont
			{Helfrich}},\ }\href@noop {} {\bibfield  {journal} {\bibinfo  {journal} {Z.
				Naturforsch.}\ }\textbf {\bibinfo {volume} {28c}},\ \bibinfo {pages} {693}
		(\bibinfo {year} {1973})}\BibitemShut {NoStop}%
	\bibitem [{\citenamefont {Kaoui}\ \emph {et~al.}(2008)\citenamefont {Kaoui},
		\citenamefont {Ristow}, \citenamefont {Cantat}, \citenamefont {Misbah},\ and\
		\citenamefont {Zimmermann}}]{kaoui2008migration}%
	\BibitemOpen
	\bibfield  {author} {\bibinfo {author} {\bibfnamefont {B.}~\bibnamefont
			{Kaoui}}, \bibinfo {author} {\bibfnamefont {G.~H.}\ \bibnamefont {Ristow}},
		\bibinfo {author} {\bibfnamefont {I.}~\bibnamefont {Cantat}}, \bibinfo
		{author} {\bibfnamefont {C.}~\bibnamefont {Misbah}}, \ and\ \bibinfo {author}
		{\bibfnamefont {W.}~\bibnamefont {Zimmermann}},\ }\href {\doibase
		10.1103/PhysRevE.77.021903} {\bibfield  {journal} {\bibinfo  {journal} {Phys.
				Rev. E}\ }\textbf {\bibinfo {volume} {77}},\ \bibinfo {pages} {021903}
		(\bibinfo {year} {2008})}\BibitemShut {NoStop}%
	\bibitem [{\citenamefont {Canham}(1970)}]{canham1970minimum}%
	\BibitemOpen
	\bibfield  {author} {\bibinfo {author} {\bibfnamefont {P.}~\bibnamefont
			{Canham}},\ }\href@noop {} {\bibfield  {journal} {\bibinfo  {journal}
			{Journal of Theoretical Biology}\ }\textbf {\bibinfo {volume} {26}},\
		\bibinfo {pages} {61} (\bibinfo {year} {1970})}\BibitemShut {NoStop}%
	\bibitem [{\citenamefont {Pozrikidis}(1992)}]{pozrikidis1992boundary}%
	\BibitemOpen
	\bibfield  {author} {\bibinfo {author} {\bibfnamefont {C.}~\bibnamefont
			{Pozrikidis}},\ }\href {http://books.google.de/books?id=qXt5bOqDEgQC} {\emph
		{\bibinfo {title} {Boundary Integral and Singularity Methods for Linearized
				Viscous Flow}}},\ Cambridge Texts in Applied Mathematics\ (\bibinfo
	{publisher} {Cambridge University Press},\ \bibinfo {year}
	{1992})\BibitemShut {NoStop}%
	\bibitem [{\citenamefont {Kaoui}\ \emph {et~al.}(2011)\citenamefont {Kaoui},
		\citenamefont {Tahiri}, \citenamefont {Biben}, \citenamefont {Ez-Zahraouy},
		\citenamefont {Benyoussef}, \citenamefont {Biros},\ and\ \citenamefont
		{Misbah}}]{kaoui2011complexity}%
	\BibitemOpen
	\bibfield  {author} {\bibinfo {author} {\bibfnamefont {B.}~\bibnamefont
			{Kaoui}}, \bibinfo {author} {\bibfnamefont {N.}~\bibnamefont {Tahiri}},
		\bibinfo {author} {\bibfnamefont {T.}~\bibnamefont {Biben}}, \bibinfo
		{author} {\bibfnamefont {H.}~\bibnamefont {Ez-Zahraouy}}, \bibinfo {author}
		{\bibfnamefont {A.}~\bibnamefont {Benyoussef}}, \bibinfo {author}
		{\bibfnamefont {G.}~\bibnamefont {Biros}}, \ and\ \bibinfo {author}
		{\bibfnamefont {C.}~\bibnamefont {Misbah}},\ }\href@noop {} {\bibfield
		{journal} {\bibinfo  {journal} {Physical Review E}\ }\textbf {\bibinfo
			{volume} {84}},\ \bibinfo {pages} {041906} (\bibinfo {year}
		{2011})}\BibitemShut {NoStop}%
	\bibitem [{\citenamefont {Hochmuth}\ \emph {et~al.}(1979)\citenamefont
		{Hochmuth}, \citenamefont {Worthy},\ and\ \citenamefont
		{Evans}}]{hochmuth1979red}%
	\BibitemOpen
	\bibfield  {author} {\bibinfo {author} {\bibfnamefont {R.}~\bibnamefont
			{Hochmuth}}, \bibinfo {author} {\bibfnamefont {P.}~\bibnamefont {Worthy}}, \
		and\ \bibinfo {author} {\bibfnamefont {E.}~\bibnamefont {Evans}},\
	}\href@noop {} {\bibfield  {journal} {\bibinfo  {journal} {Biophysical
				journal}\ }\textbf {\bibinfo {volume} {26}},\ \bibinfo {pages} {101}
		(\bibinfo {year} {1979})}\BibitemShut {NoStop}%
	\bibitem [{\citenamefont {Tomaiuolo}\ and\ \citenamefont
		{Guido}(2011)}]{tomaiuolo2011start}%
	\BibitemOpen
	\bibfield  {author} {\bibinfo {author} {\bibfnamefont {G.}~\bibnamefont
			{Tomaiuolo}}\ and\ \bibinfo {author} {\bibfnamefont {S.}~\bibnamefont
			{Guido}},\ }\href@noop {} {\bibfield  {journal} {\bibinfo  {journal}
			{Microvascular research}\ }\textbf {\bibinfo {volume} {82}},\ \bibinfo
		{pages} {35} (\bibinfo {year} {2011})}\BibitemShut {NoStop}%
	\bibitem [{\citenamefont {Prado}\ \emph {et~al.}(2015)\citenamefont {Prado},
		\citenamefont {Farutin}, \citenamefont {Misbah},\ and\ \citenamefont
		{Bureau}}]{prado2015viscoelastic}%
	\BibitemOpen
	\bibfield  {author} {\bibinfo {author} {\bibfnamefont {G.}~\bibnamefont
			{Prado}}, \bibinfo {author} {\bibfnamefont {A.}~\bibnamefont {Farutin}},
		\bibinfo {author} {\bibfnamefont {C.}~\bibnamefont {Misbah}}, \ and\ \bibinfo
		{author} {\bibfnamefont {L.}~\bibnamefont {Bureau}},\ }\href@noop {}
	{\bibfield  {journal} {\bibinfo  {journal} {Biophysical journal}\ }\textbf
		{\bibinfo {volume} {108}},\ \bibinfo {pages} {2126} (\bibinfo {year}
		{2015})}\BibitemShut {NoStop}%
	\bibitem [{\citenamefont {Ghigliotti}\ \emph {et~al.}(2010)\citenamefont
		{Ghigliotti}, \citenamefont {Biben},\ and\ \citenamefont
		{Misbah}}]{ghigliotti2010rheology}%
	\BibitemOpen
	\bibfield  {author} {\bibinfo {author} {\bibfnamefont {G.}~\bibnamefont
			{Ghigliotti}}, \bibinfo {author} {\bibfnamefont {T.}~\bibnamefont {Biben}}, \
		and\ \bibinfo {author} {\bibfnamefont {C.}~\bibnamefont {Misbah}},\
	}\href@noop {} {\bibfield  {journal} {\bibinfo  {journal} {Journal of Fluid
				Mechanics}\ }\textbf {\bibinfo {volume} {653}},\ \bibinfo {pages} {489}
		(\bibinfo {year} {2010})}\BibitemShut {NoStop}%
\end{thebibliography}

\end{document}